\documentclass{book}
\usepackage[utf8]{inputenc}
\usepackage{xcolor}
\usepackage{epigraph}
\usepackage{graphicx, subfig}
\usepackage[sort&compress,comma,numbers]{natbib}
\usepackage{xurl}
\usepackage{hyperref}
\usepackage{caption}
\begin{document}

\setcounter{chapter}{22}  
\chapter{Superspreading and Heterogeneity in Epidemics}
\chaptermark{{\scriptsize Diffusive Spreading in Nature.., A. Bunde et al, Springer}}
\subsection*{Klaus Kroy}
\epigraph{C’est une id\'ee qui peut faire rire, mais la seule fa\c con de lutter contre la peste, c’est l’honn\^etet\'e. }{Albert Camus, \emph{La Peste}}

\section{Introduction}  
Molecules and colloids spread through fluids, genes spread through species, idioms through vernacular communities, and fads and pathogens  through social networks\dots That erratic dynamics on a micro-scale gives rise to  universal systematic  spreading patterns on a macro-scale is a ubiquitous observation, transcending  scientific disciplines. Following Fourier and Fick, \emph{the stochastic spreading of heat and small solutes is macroscopically well described by a deterministic, hydrodynamic law},  namely the diffusion equation. It provides formidable forecasts and technical control (see, e.g., the introductory Chapter 2 of this book). And it can serve as a template for more complex spreading processes: from the propagation of quantum wave functions in imaginary time to sub- and super-diffusion  in viscoelastic and  ``active’’ matter, or genetic drift and stock-option pricing \cite{frey-kroy:2005}. But how far does it take us in the attempt to quantify and forecast  epidemic pathogen and disease spreading?

This question is deemed of such importance that a whole discipline of applied science is devoted to it: epidemiology. Pathogens spread not only within, between, and among but also \emph{with} their generally mobile host organisms. Therefore the processes of interest in epidemiology could suggestively be characterized as ``piggyback spreading'' or ``spreading squared''. The phenomenon shares many similarities with the other spreading processes mentioned above. One is that researchers feel tempted to model it by diffusion. Indeed, classical epidemiological theories routinely describe disease spreading in the spirit of chemical reactions in a test tube. And, as reviewed by Dirk Brockmann in Chapter~20, they can even schematically account for non-trivial human mobility patterns if formulated on a warped space. But there are also some more erratic elements of epidemic disease spreading, reminiscent of more unwieldy phenomena such as earthquakes, hurricanes, traffic jams, and stock-market crashes. They are  characterized by extreme heterogeneities, coupled across diverse scales. Rare chains of micro-scale fluctuations may occasionally get heftily magnified into large erratic outbreaks. One then speaks of a ``high tail risks'' due to an uncomfortable combination of the large size and impact of an expected event with its estimated small likelihood. Such traits are in sharp contrast to what ordinary diffusion and classical epidemiological models would predict and make the forecasting and control of epidemics  intrinsically and notoriously hard.  

Therefore, the aim of the present chapter is to provide a (non-expert\footnote{by a soft-matter physicist looking at this multifaceted field from the outside}) synopsis of the  crucial role of  \emph{spatial, probabilistic, and genetic heterogeneity in real-world epidemics}.  Their importance for the failure of conventional epidemiological models was already anticipated by some of the pioneers in the field~\cite{heesterbeek:2005}. The ``false analogy between infection in disease and the mechanism understood under the name of chemical mass action'' (H.~E. Soper, 1929) was criticized, and first steps were taken (notably by A.~G. McKendrick) to resolve the issue. Today, we can draw on  cross-fertilizations from network and game theory and the emerging field of eco-evolutionary dynamics to substantiate such worries.   

The notion of \emph{superspreading} can serve as a guiding theme. In a narrow sense, it refers to the strong  variability in the number  of secondary infections or ``offspring'' of a contagious pathogen carrier \cite{LloydSmithPaper}. This probabilistic heterogeneity or so-called overdispersion (compared to an idealized test-tube reaction) means that the majority of all infections is not caused by typical encounters but by a small number of exceptional superspreading events. These are intuitively expected to be caused by ``supercarriers'' and ``supershedders'' of pathogens.\footnote{Indeed, a large recent  study investigating superspreading in the COVID-19 epidemic found that 2\% of positively tested individuals carry 90\% of the circulating virus  \cite{2PercentCarry90Percent}.} But,  as further elucidated below, one should not rashly jump to such conclusions. I want to use the term in a wider sense, also comprising a heterogeneous susceptibility as well as the heterogeneity of the pathogen carriers' social contact networks and  spatio-temporal dynamics. Furthermore,  it makes sense to also implicate an extended hierarchy of superimposed levels of spreading, similar to the discussion in Chapter~19 of this volume. Namely, beyond the mentioned multi-layered ``spreading squared'' of  pathogens hitchhiking with their mobile hosts (and possibly intermediate ``vectors'', such as mosquitoes or other transmitting animals), there are further, less palpable, but similarly heterogeneous spreading processes involved. Think of mutation patterns  spreading along a pathogen's genealogical or phylogenetic tree, for example. Thereby, epidemic dynamics is always confounded by population genetics~\cite{anderson-may:1991}. And, as discussed below, one may even ask, whether ``genetic drift'' (i.e., diffusion in sequence space) may explain ``superspreading events without superspreaders''~\cite{SuperspreadingWithoutSuperspreaders,vignuzzi-etal:2006}? Finally, there is the  overlaid spreading of information about the course of an unfolding epidemic, which may itself ``go viral'' and thereby induce highly nonlinear feedback effects~\cite{DynamicalInterplayAwarenessEpidemicSpreadingNetworks}.
Altogether, epidemics thus have to be regarded as a manifestation of a (quite fickle) \emph{complex dynamical system}\footnote{\label{ftn:schrappe}M Schrappe \emph{et al.}, 2021: ``Thesenpapier 8.0 zur Pandemie durch SARS-CoV-2/Covid-19'' \url{http://doi.org/10.24945/MVF.05.21.1866-0533.2348}; \& Addendum: \url{http://www.monitor-versorgungsforschung.de/efirst/Mueller_Addendum_Thesenpapier-8-0_Modellierung}}. While this may seem obvious to some experts, it is not always widely appreciated.   

\section{Microbes, military, and malady}
Speaking of pathogen and disease spreading,  a few facts about bacteria and viruses seem in place. First of all, it is important to realize that they are integral parts of any ecosystem. And that the whole biosphere relies on (and is dominated by) microorganisms that have for the largest part not yet been scientifically investigated. 
Even within our own bodies, bacteria and viruses by far outnumber ``our own" cells.  
In particular, viruses are the most abundant and fastest mutating genetic elements on Earth --- their number of offspring in a single infection event easily exceeds the size of the whole host population. And, although not counted among the proper lifeforms, they play a decisive role in shaping the tree of life, as ``agents of evolutionary invention''~\cite{cordingley_2017}. By infecting us, or more often the broad array of microorganisms inhabiting us, they mould our genetic and transcriptional identity. 
Even the small minority among them identified as pathogenic usually does little or no harm in their natural reservoir species.  But it may occasionally cause significant disease outbreaks at overpopulated gatherings of hosts, or when jumping between cohabiting species. 

This has become a serious issue with the onset of urbanisation and intensive farming in the wake of the neolithic revolution (see Chapters 16 and 17 of this volume), which has culminated in industrial poultry farms~\cite{HerdImmunPoultry}, care and nursing homes~\cite{ForProfitLongTermCareHomesRiskCov19}, 
 and crowded sickbays, prisons~\cite{leibowitz-etal:2021}, and warships~\cite{USNavyInfluOut}. 
Indeed, all types of herding and crowding are closely interwoven with infectious diseases. This applies particularly to warfare, where disease has often trumped combat by its death toll~\cite{war-epidemics:2004,NextGenBioweapons}: from the epidemic ``killing the warriors in droves” evoked by Homer to spark the story of his Iliad, to the plague of Athens in 430 BC that extinguished about a quarter of those entrenched in the besieged overcrowded city, and the Spanish flu pandemic, with (far) more than 20 million victims, during the First World War. 

\begin{figure}
    \centering
    \includegraphics[width=\textwidth]{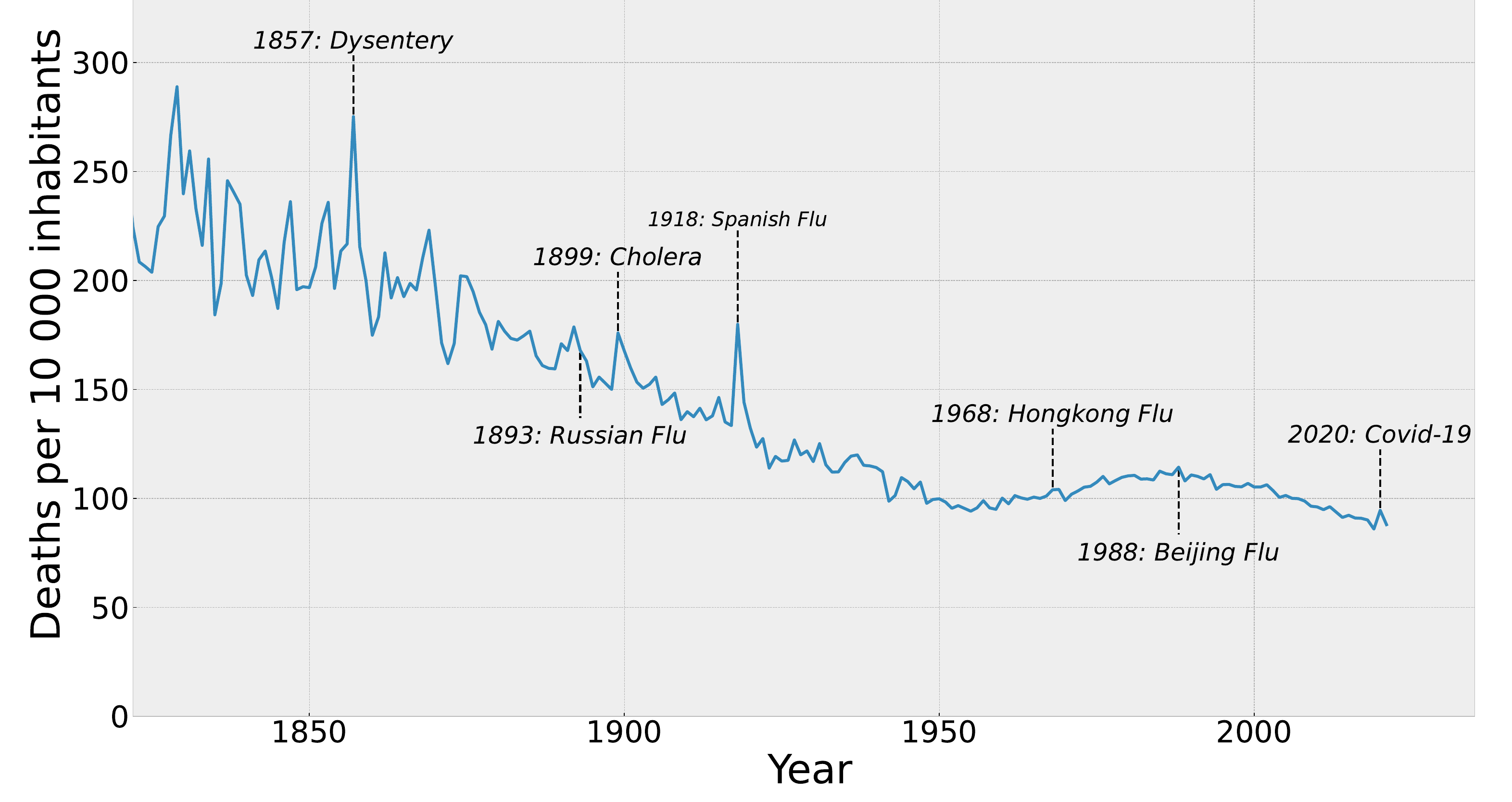}
    \caption{\textbf{Impact of historical epidemics.} The annual number of deaths per ten thousand inhabitants in Sweden throughout the last centuries (up to the year 2021) exhibits signatures of major epidemics piercing an overall trend of increasing  longevity (Statistiska Centralbyrån$^{\ref{ftn:swede}}$).}
    \label{fig:sweden}
\end{figure}

Unsurprisingly, military forces have therefore become leading proponents of germ and vaccine research. George Washington’s order to variolate the entire American Army against smallpox is sometimes cited as decisive for the American civil war \cite{USVaccineDevelopment}. The success story could however not be repeated by the enthusiastically advocated  vaccination campaigns, targeting epidemiologically irrelevant bacteria such as Pfeiffer's ``bacillus influenzae'', against the Spanish flu~\cite{eyler:2010}. From the ensuing controversies emerged an early set of criteria for valid vaccine trials, while the types of viruses now held responsible for transmitting the flu, as well as their immune and vaccine escape capabilities, remain mysterious and challenging,  another century later~\cite{suzuki:2006,domingo-perales:2019,bolton-etal:2021}.
Albeit, till today, the military remains a central player and an incubator for vaccinologists and epidemiologists, many of which later move on to industry and academia, contributing to notable discoveries, e.g., of vector-borne diseases and transmissions, the oral polio vaccine, etc. Together with increased prosperity and improved sanitation, 
they are widely credited for taming the hazardous impact of contagious diseases on our highly interconnected modern societies  (Fig.~\ref{fig:sweden})\footnote{\label{ftn:swede}\url{http://www.scb.se/en/finding-statistics/statistics-by-subject-area/population/population-composition/population-statistics/pong/tables-and-graphs/yearly-statistics--the-whole-country/population-and-population-changes}}.
On a less optimistic note, such activity has unavoidably been entangled with bioweapon development\footnote{As remarked by an insider, ``everything is dual use --- the people, the facilities and the equipment'', S. Husseini, Independent Science News 2020: \url{http://www.independentsciencenews.org/news/peter-daszaks-ecohealth-alliance-has-hidden-almost-40-million-in-pentagon-funding}. See also S. Lerner, M. Hvistendahl, M. Hibbett, The Intercept 2021:
 \url{http://theintercept.com/2021/09/09/covid-origins-gain-of-function-research}}, (at least) ever since “variolated’’ sundries, blankets, and handkerchieves were distributed around Fort Pitt in Pennsylvania to decimate the Native Americans, in the 1760’s~\cite{USVaccineDevelopment}.  And although the ensuing “gain-of-function”\footnote{For an introduction see: \url{http://www.youtube.com/watch?v=zVOaXdKc-DI} or \url{http://www.youtube.com/watch?v=Q0nuyPQzU18}} research has become increasingly unpopular and repeatedly officially banned~\cite{lipsitch:2018}, it has arguably never ceased~\cite{NextGenBioweapons}. The late Steven Hawking's worries\footnote{S. Hawking: “we face a number of threats to our survival, from nuclear war, catastrophic global warming, and genetically-engineered viruses” in his Reith Lecture 2016: ``Black holes ain't as black as they are painted'', \url{http://www.youtube.com/watch?v=ljvVPAZHnD4}} about disasters more imminent than those predicted by his cosmological theories thus do not seem to have arisen out of thin air, but rather underscore a commonly underestimated eminent political dimension of epidemiology that has resurfaced in recent debates~\cite{ArisenViaSerialPassage, UnravelOriginsCovid19,InvestigateOriginsCovid19}.

\section{SIR-storm in a teacup}\label{sec:sir}
The classical   ``test-tube'' models of epidemiology \cite{math-epidemiology:2019}, based on seminal work by Kermack and McKendrick~\cite{kermack-mckenrdrick:1927}, were briefly introduced in Chapters 20 and 21 of this volume and, more thoroughly,  by the applied mathematician Stephan Luckhaus in his Leipzig  lectures.\footnote{\label{ftn:lucky}S. Luckhaus, 2020 recordings: \url{http://www.youtube.com/watch?v=SZ4dIEb2ttM} (English), \url{http://www.youtube.com/watch?v=FJYjqltJn9E} (German); see also: \url{http://www.mis.mpg.de/preprints/2020/preprint2020_105.pdf}} They are better known as compartmental models, since they describe the mutual ``chemical reactions'' of idealized homogeneous fractions or compartments of a given population, such as the \textbf{S}usceptible, \textbf{I}nfected, \textbf{R}ecovered, etc. In their simplest form, these models completely disregard the spatial, social, and biological heterogeneity of the population, emphasized in the present contribution. Although one therefore cannot trust their numerical results, they make a number of \emph{qualitatively robust predictions} that are worthwhile recalling.

A major phenomenological observation  that is captured well by the standard SIR-type models is the characteristic time course of epidemics.  They proceed via overshooting \emph{infection waves} toward 
\emph{i)} extinction due to herd immunity or \emph{ii)} an endemic state, respectively. The first scenario \emph{i)} applies to a ``static'' host population and pathogen, for which the (nominal) herd-immunity threshold follows from the fraction of remaining susceptible individuals at peak infection. The second, more complex, scenario \emph{ii)}  takes into account the gradual emergence of new susceptible individuals, e.g., due to pathogen mutations, (reverse) zoonosis\footnote{For example, SARS-CoV-2 has remarkably rapidly established so-called ``animal reservoirs'', e.g., in cat, dog, tiger, lion, puma, mink, and white-tailed deer~\cite{BatsPangolinsMinksOtherAnimals,WildWhiteTailedDeer}, which was  answered by mass cullings of tens of millions of minks ``to protect the vaccines''~\cite{MassCullingOfMinks}}, waning immunity, and/or the host population's turnover (e.g.\ by births and migration). These processes slowly and inconspicuously shift the balance in favor of new outbreaks, thus leading to repeated infection waves.   Irrespective of the precise scenario, a robust conclusion of the standard models is that only a fraction of the population ever becomes infected during an infection wave.\footnote{In the first  COVID-19 infection wave,  about 0.2\% of the German population tested PCR-positive. Almost two years and several waves and mutants later, everybody had statistically been PCR-tested at least once, yet  only about 6.5\% positive. Such low infection rates hint at substantial population heterogeneities and cross-immunities (epidemiological ``dark matter'')} 

Conceptually, the models are furthermore helpful for  sorting the reaction rates into two classes: those setting the speed of spreading and the late-time endemic state or herd immunity, respectively.  The distinction mirrors that 
between Arrhenius- and Boltzmann-factors in physical chemistry: the former only modulate the speed of a chemical reaction, while the latter alone determine its outcome via the mass-action law.\footnote{This equilibrium-type constraint confers a crucial element of predictability to epidemics, \emph{if their population-wide pathogen-host chemistry is known and remains fixed over time}} The fundamental insight for pathogens that cannot be eradicated is that mitigation measures, including face masks, non-sterilizing vaccines, and pharmaceuticals, are mere (anti-)catalysts. They can at best~\cite{ioannidis:npj2021} \emph{affect the time course but not the long-term fate} of an epidemic, which is determined by the pathogen-host chemistry, alone. For example, rapid border closures may help to gain some time during the early stage of an epidemic. Yet, even the draconian travel quarantine imposed in Wuhan during the onset of the COVID epidemic in early 2020 was estimated to delay the further progression in Mainland China by no more than 3 to 5 days~\cite{EffectTravelRestrictions}. And even island countries such as Australia, New Zealand, Taiwan, and Iceland, where infection numbers were kept extremely low until the vaccination, anything but ``almost avoided the pandemic entirely''\footnote{\url{https://www.uschamber.com/on-demand/coronavirus-pandemic/bill-melinda-gates-on-the-pandemic-and-what-comes-next?autoplay=1} (min 6:28).}, thus ``vindicating Boltzmann over Arrhenius''. Such sobering insights can temper expectations about what pharmaceutical and non-pharmaceutical interventions can generally achieve for the control of epidemics.  As the examples suggest, social interventions are notorious for their hard to assess~\cite{chin-etal:2021} and hard to justify cost-benefit ratios among the experts~\cite{BiosecurityDiseaseMitigationMeasures, LockdownReportAllen, coccia:2021,milan-etal:2021}.  Despite some initial empirical~\cite{DidLockdownsServeTheirPurpose}, but more often lopsidedly  model-based\footnote{\label{ftn:lobs}P. W. Magness, AIER 2021: \url{http://www.aier.org/article/the-failure-of-imperial-college-modeling-is-far-worse-than-we-knew}; G. A. Quinn \emph{et al.}, 2021: \url{http://doi.org/10.31219/osf.io/s9z2p}}, propositions to the contrary, also the latest epidemic does not seem to provide a plausible exception to this  rule~\cite{gibson:2020,AssessingMandatoryStayAtHomeEffectsOnSpread, RethinkingLockdownGroupthink,watanabe-tomoyoshi:2021,kuhbandner-etal:2022,herby-etal:2022}. 

\section{Transmission heterogeneity}
The perplexingly unimpressive effect of global ``bulk'' interventions on epidemic spreading clearly begs for further explanation. A first clue can be gained if the compartmental models are extended to account for some population heterogeneity by dividing the population up into cohorts with distinct properties. They then admit richer solutions, with some variability how the overall infection rate is distributed among the cohorts.  It is then  intuitively clear that mitigation measures  can be more effective and more economic if they target specific cohorts rather than the whole population~\cite{anderson-may:1991,diekmann:2000}.  Epidemiologists have long known how to exploit the effect to protect particularly vulnerable groups$.^{\ref{ftn:lucky}}$ COVID-19 again provides a good example. Only a small percentage of the elderly population with comorbidities is at serious risk and accounts for almost all deaths~\cite{AgeRelatedSevereOutcomes, AgeSpecificMortalityImmunityPatterns}, initially about half of them in long-term care-homes~\cite{HighImpactCov19LongTermCareFacilities}. According to the extended compartmental models with multiple cohorts, the  conventional mitigation strategy is then to let the disease spread quickly\footnote{At which point a long-term investment in the healthcare sector appears to pay its way~\cite{coccia:2021}} among the non-vulnerable  (the young and healthy~\cite{PreactivatedAntiviralInnateImmunityChildren,borch-etal:2022}), while the vulnerable group temporarily self-isolates. Altogether, the extent of the infection wave in the vulnerable group---and, in a well mixed population, even their herd immunity threshold---can thereby be reduced, resulting in a much lower overall cumulative death toll.  The important take-home message is that it is crucial to account for the population heterogeneity, with respect to both vulnerability and social contacts. Well-intentioned indiscriminate or even misaligned interventions (e.g., off-season for a seasonal disease or targeting the less instead of the more vulnerable cohorts) may be counter-productive, delaying the necessary immunization and resulting in an overall increased death toll, at the end of the line. Curiously, experts who voiced such textbook wisdom during the early days of the recent COVID epidemic encountered fierce political opposition\footnote{\label{ftn:censor}J. Levin, New York Post 2020: \url{http://nypost.com/2020/05/16/youtube-censors-epidemiologist-knut-wittkowski-for-opposing-lockdown/};  M. Kulldorff, The Spectator 2021: \url{http://www.spectator.co.uk/article/covid-lockdown-and-the-retreat-of-scientific-debate}; S. Luckhaus \#wissenschaftstehtauf 2021: \url{http://odysee.com/@wissenschaftstehtauf:8/Luckhaus_Bauchbinde_final:b};  The Wall Street Journal Editorial Board 2021: \url{http://www.wsj.com/articles/fauci-collins-emails-great-barrington-declaration-covid-pandemic-lockdown-11640129116}}.  

A number of recent studies~(see Ref.~\cite{neipel-etal:2020} and therein) have investigated extensions of the cohort approach to  broad distributions of the susceptibility, in order to account for the substantial diversity in the transmission probabilities. Jülicher and coworkers considered a power-law distribution of (natural) susceptibilities, with its exponent 
$\alpha$ being the only new parameter added to the standard SIR model~\cite{neipel-etal:2020}. Key properties of the epidemic wave dynamics, including the infection curve, the herd immunity level, and the final size of the epidemic, can then still be calculated exactly, while avoiding the proliferation of poorly defined model parameters that generally plagues the modeling of heterogeneities in terms of several discrete cohorts.  An important finding is that the average dynamics is naturally sub-exponential, $\alpha$-dependent, and allows for \emph{herd-immunity thresholds as low as a few per cent} of the total population.  The model also predicts how the transmission probability distribution itself evolves dynamically throughout an epidemic, as the numbers in the susceptible cohorts change and eventually converge to an invariant distribution $\propto x^{\alpha-1}e^{-\alpha x}$.  

While this is certainly an important step towards reality, further modifications of the test-tube caricature of epidemic spreading become necessary if one wants to account more thoroughly for the whole spectrum of relevant heterogeneities~\cite{EpidemicProcessesComplexNetworks} and include, for example, an uneven spatial or social mixing within or between the cohorts, as outlined in the following. 

\section{Heterogeneous contact networks}
Heterogeneous populations should of course not be expected to be well mixed. Their contacts are unevenly distributed with respect to their number as well as to other parameters such as age group and profession. Referring back to the schematic example of the two cohorts, one will only reap the full benefit of a targeted protection strategy if the population is well mixed. If the vulnerable remain permanently segregated, instead, in close mutual contact rather than dispersed within an already largely immunized population, the risk of devastating outbreaks among them is not diminished as much as it could be (``dry-tinder effect'')\footnote{N. Rondinone, CT Insider 2021: \label{nursing}\url{http://www.ctinsider.com/news/article/What-we-know-about-one-of-CT-s-deadliest-16640827.php}}. Generally speaking, beyond a population's susceptibility distribution, one also needs to address its heterogeneous contact networks, if one wants to fully understand the spreading of an epidemic. In \emph{clustered networks}, the connections are skewed to favor members of the same clique. Cliques are densely intra-connected but only weakly inter-connected. Accordingly,  in such networks, infections initially  spread very well but then slow down and remain more confined than naively expected. In other words, the fire of infection burns fast but extinguishes easily. To reach the next clique or cluster, the epidemic has to pass via narrow population \emph{bottlenecks}, where only a few pathogens can pass, so that details matter and chance rules~\cite{SmerlakBottleneck}. (This ``reign of small numbers''~\cite{frey-kroy:2005} plays a prominent role in virus evolution, as discussed further below.) 

The topological feature of contact clustering may thus considerably impede the global spread of a disease compared to  tree-like contact networks, while it facilitates contact tracing~\cite{PropertiesHighlyClusteredNetworks,eames:2008}. Besides nursing homes, typical real-world examples for clusters could be households and school classes, say, but also an infected organism or its organs. While a pathogen may find it easy to spread within one organ, it may turn out difficult to invade the next, and similarly for social groups. As a consequence, real-world epidemics only reach a smaller part of the population than predicted by test-tube models, yet most of it relatively easily. This conclusion resonates well with the above-mentioned findings by Jülicher and coworkers. And indeed, their approach can also be interpreted as an attempt to effectively incorporate some contact heterogeneity into the simple SIR model. In practice, however,  clustering and distributed susceptibility or transmissibility are distinct features that interfere with each other \cite{grossmann-etal:2021}. 

To make things worse, they interfere with yet another non-trivial but common topological trait of natural contact networks, namely their  heterogeneity with respect to the number of contacts emanating from a network node (its ``degree''). In his book ``Linked -- The New Science of Networks'' \cite{LinkedBarab},  A.-L. Barabasi emphasizes this other crucial aspect of real-world contact networks, known as their \emph{broad degree distribution}. He vividly recounts
the story of one of the first known superspreaders, often called Patient Zero of the AIDS epidemic. Who turned out to be a central hub (a particularly highly connected, i.e.~high-degree, network node), with an estimated number of 2500 links or more, in a large network of sexual contacts. At least some 40 early patients could explicitly be traced back to him. This is a manifestation of the empirical observation that social and sexual contact networks are very heterogeneous~\cite{WebSexContact}. The weakly connected nodes (with zero or one sex partner, say) are most frequent, but the number of more highly connected hubs decays only slowly, typically like a power-law, as a function of their connectivity (their degree). This means that individuals will not spread a disease simply  proportionally to their pathogen load.  Indeed, most do not spread a pathogen at all, but a few a lot\footnote{Not unexpectedly, staff members feature prominently in the timeline of the well documented nursing-home outbreak referenced in footnote~\ref{nursing}.}. Yet, neither do all hubs in a network contribute simply proportionally to their degree, not even if one artificially assumes  homogeneous pathogen load and transmission. And it would not be easy to localize them all, anyway, or even to define and select the structurally most relevant ones~\cite{IdentificationOfInfluentialSpreaders, UnificationApproachesEpidemic}.  Unsurprisingly, the best search tool known to solve this complex task is actually an epidemic, which may in this context be thought of as a ``swarm algorithm''.

\begin{figure}
        \centering
    \includegraphics[width=\textwidth]{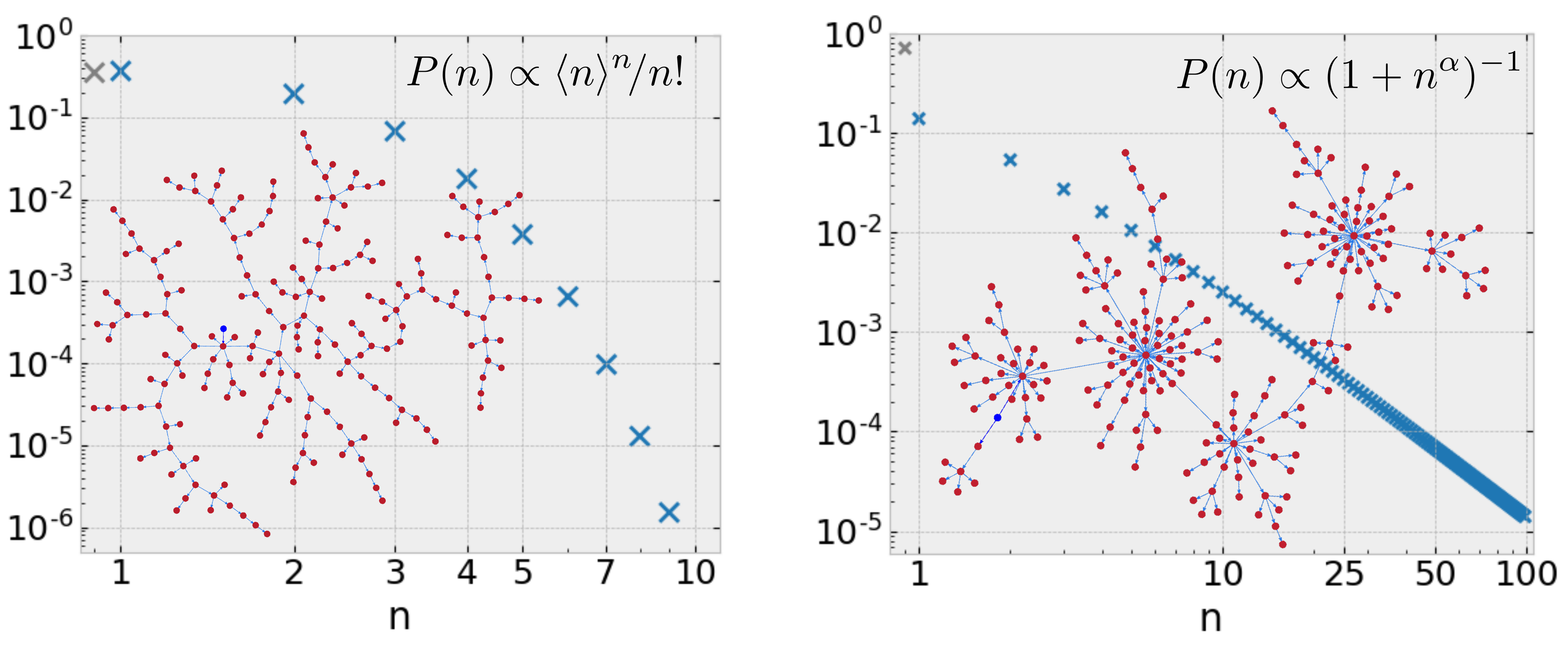}
     \caption{\textbf{Homogeneous versus heterogeneous spreading.} Transmission trees grown from a single seed with offspring numbers from a Poisson distribution (left) or a truncated power-law (right, $\alpha\approx2.35$) with gray \textcolor{gray}{$\times$} indicating $P(0)$. In the Poisson or test-tube setting, most infected nodes pass on a spreading disease: average behavior matters. The power-law case (also overdispersed, fat-tailed, scale-free) is more realistic for human contact networks~\cite{EpidemicProcessesComplexNetworks,WebSexContact}. Its (identical) average hardly matters,  as most nodes infect no others, and the epidemic is driven by rare, explosive but unstable \emph{bursts}~\cite{barabasi:bursts} or superspreading events. The more fat-tailed $P(n)$, ``the more the tail wags the dog''~\cite{TailRiskDiseases,rudiger-etal:2021}.} 
    \label{fig:inftree}
\end{figure}

Similarly as for the broadly distributed transmissibilities, discussed further above, an important and robust epidemiological consequence of broadly  distributed contact numbers is the lowering of the epidemic threshold \cite{InfectionDynamicsScaleFreeNetworks}. In the standard SIR models, which describe relations between population averages, the fate of an epidemic is determined by its  effective reproduction number $R$ of secondary infections caused, on average, by an infected individual. The epidemic is predicted to spread exponentially for $R>1$ and to decay exponentially for $R<1$. However, for the important class of scale-free networks (with power-law degree distributions), the epidemic threshold is \emph{formally} found to vanish in an infinite network: in strongly heterogeneous networks, pathogens (no matter how weakly contagious) will always spread to a certain macroscopic scale. And they do so in a rather heterogeneous fashion, as illustrated in Fig.~\ref{fig:inftree}.  With increasing broadening of the degree distribution, occasional disease extinction becomes more likely and outbreaks rarer but more explosive --- and all this even without invoking the non-trivial effects due to distributed transmissibilities, clustering, and bottlenecks. In other words, \emph{real-world epidemic spreading is, for more than one reason, prone to generating considerable tail risks}.

\section{A tale of tail risks}
To gain some intuition for the tail-risk effects associated with power-law distributions, consider the following gamble. A banker tosses a fair coin for you until tail comes up. What is the fair price for a round, if you earn $2^{n+1}$ (in your favorite currency) for $n$ heads in a row before tail comes up? The problem with this game, and similar real-word processes in finance and epidemics, is that the chance $2^{-n-1}$ for the row of $n$ heads is the reciprocal of the profit $2^{n+1}$. So the probability distribution for earning $x_n=2^{n+1}$ is given by the power law $x_n^{-1}$. The most striking consequence of this is that your most likely payoff is 2 while the average payoff $\sum_n x_n/x_n=\sum_n 1$ diverges. Accordingly, your likely gain in a single round is fairly small, but the ``fair'' price is infinity. Notice the so-called Petersburg paradox associated with this result. Namely, that a banker charging a better-than-fair (any finite) price per round, say 1000, can play the game for a long time and accumulate impressive wealth,  before he eventually goes bankrupt and needs to be bailed out by the tax payers. 

Beyond illustrating the key problem with power-law risks, which indeed seem to arise ubiquitously in non-equilibrium real-world  processes from plate tectonics to financial markets~\cite{kiyoshi-sornette:2021}, the toy model is also instructive in another sense. It shows that high tail risks do not always require very heterogeneous conditions as a necessary prerequisite. They can even result from statistically independent random events, like coin tosses, if the process is capable of producing exponential growth.  Which leads back to epidemiology, where the gambler's profit amounts to pathogen proliferation, of course, and the banker's allegedly 100\% safe business model is the analogue of the epidemiologist's  confident prediction of the next infection wave. Since, according to the standard models, epidemics are of course the textbook paradigm for exponential growth. Or are they?

\section{Spatial spreading}
So far, the discussion of heterogeneities has focussed on topological features. However, unlike fads or idioms, contagious diseases do not spread via phone calls and emails but require physical encounters. Which means that the mentioned heterogeneous social networks need to be considered together with their embedding into real space.
And the physical geometry of the latter, as encoded in a city map or by  regions of inhabitable land on the globe, brings new heterogeneities into the equation that can potentially override the mentioned topological features or even change them~\cite{west:scale}. For example, the topological feature of clustering, discussed above, may to some extent be thought to arise as consequence of spatial embedding (to households, work places, \dots). Also, so-called superspreading events, with an explosive growth of infections, clearly seem to be tied to  specific patterns of spatial crowding and mobility: besides the mentioned nursing homes and warships also large conferences~\cite{BostonSuperspreading},  cold slaughter houses~\cite{guenther-etal:2020}, and hot parties~\cite{brandal-etal:2021,wessendorf:2021}  (with poor air conditioning) --- but not public trains~\cite{LongDistanceTrains}, the exchange of bank notes~\cite{ECB:2021}, or outdoor events~\cite{bulfone-etal:2020}.  Once more, a small fraction of (in this case spatial\footnote{\label{ftn:3C}The so-called three C's: closed spaces, crowded places, close-contact-settings}) interaction patterns appears to produce almost all spreading~\cite{chang-etal:2021}. 

A crucial role of the spatial geometry has also been established for the process of gene spreading, which is closely related to pathogen spreading. Genomes can remain confined with their phenotypes to certain habitats for a long time, before the latter embark on larger excursions, leading to so-called founder effects. As elaborated in Johannes Krause's popular book~\cite{krause:gene}, the co-evolution of largely isolated  social groups with specific pathogens is moreover suspected to have caused major historic epidemics upon invasion of new, already inhabited territories. This is also a main topic in the history of colonialism and imperialism~\cite{watts:1997,borntodie:1998}.  In return, one may speculate that the rapid growth in human mobility and globalization throughout the last 150 years\footnote{Think of Jules Verne's bestseller ``Le Tour du monde en quatre-vingts jours'' from 1873}, could  plausibly partly explain the decaying impact of epidemics, as apparent from Fig.~\ref{fig:sweden}; namely, chiefly via the global distribution of pathogens and cross-immunities it has brought about~\cite{thompson-etal:2019}. Counter-intuitively as it may sound to some, it would mean that, in the long run, the severity of epidemics is best tamed by frequent traveling and cross-country social gatherings, and that Christmas~\cite{brandal-etal:2021}, carnival~\cite{wessendorf:2021}, and apr\`es-ski~\cite{borena-etal:2021} parties  may inadvertently serve some noble purpose. 

Not only to address such intriguing open questions, the embedding of epidemic spreading into real-space geometry is of major interest. Two key elements to consider are then the spatial heterogeneity of the population density and its heterogeneous mobility patterns~\cite{EpidemicProcessesComplexNetworks}.  A scaling approach, pioneered and popularized by  Geoffrey West~\cite{west:scale} covers both in one stroke, without necessitating the underlying mechanistic details to be resolved. It is consistent with the empirical finding that socioeconomic metrics of urban life, such as the average wages or the number of restaurants, theaters, lawyers, crimes, and cases of AIDS and flu, all scale in a more or less universal, superlinear manner with city size. This approach has also tentatively been applied to the early attack of the COVID-19 pandemic~\cite{stier-etal:2020}, and a suggestive correlation between mortality and the so-called \emph{population weighted density}\footnote{\label{ftn:weight}Swiss Policy Res.~Group, 2021: \url{http://swprs.org/judgment-day-sweden-vindicated}}  seems to hint in a similar direction.

Chapter~20 of this volume follows a more explicit route for the example of air traffic, along similar lines as pursued in invasion biology~\cite{suarez-holway-case:2001}: if you can take a plane, this may substantially affect what you (as well as the species you carry along) perceive as the metric of the globe. One can capture such heterogeneous mobility by means of a warped metric, i.e., very much in the same spirit as for diffusion in curved spacetime~\cite{DiffusionInCurvedSpacetimes} or for the case of non-isothermal Brownian motion~\cite{rings-etal:2010}, described in Chapter~8 of this volume. One can also try to take the full spectrum of means of transport into account, allowing travellers to jump the line with the help of planes, trains, busses and cabs, say. A tractable approach to do so has already been explored in the closely related context of eco-evolutionary dynamics~\cite{hallatschek-fisher:2014}. Translated to the spread of an epidemic, its predictions imply that the textbook wisdom of exponential epidemic growth requires substantial revision, as they suggest a very different (even logarithmic) intermediate asymptotic dynamics of disease spreading, under such circumstances.  Recent epidemiological studies arrive at similar conclusions, albeit on entirely different grounds~\cite{viboud-etal:2016,zonta-etal:2021}. Also the contribution by Armin Bunde and coauthors in Chapter~21 points out that the transport dynamics on a spatial network will inevitably become algebraic rather than exponential, when the network is poised towards breakup (i.e., near to its percolation threshold).  

Summarizing the above discussion, the main problem with the epidemic standard models is that the infectiousness of the infected, the susceptibility of the susceptible and recovered, and the social and spatial contact networks connecting them are in reality all far from homogeneous. This entails a substantial quantitative renormalization of the standard epidemiological model predictions --- chiefly, a drastically reduced overall herd immunity threshold and endemic pathogen concentration. But, conjointly with the propensity to (locally) exponential growth dynamics, the heterogeneities furthermore create substantial tail risks that threaten to severely undermine the mechanistic logic of the widely employed test-tube models, altogether. This problem is addressed next.

\section{Breakdown of macroscopic determinism}
The forgoing discussion has highlighted the similar and dissimilar roles of clustering, broad degree distributions, as well as biological, spatial, and mobility heterogeneities. But how all of them interfere in real-world epidemics is so far still poorly understood. And even if some effects might be hoped to compensate (rather than amplify) each other, the detailed mechanisms are far from clear, a century  after such concerns were first raised and tentatively addressed~\cite{heesterbeek:2005}. If taken literally, all of them involve problematic limiting procedures that do not peacefully coexist. For example, as mentioned above, taking a power-law degree distribution seriously for an infinite network would predict a vanishing epidemiological threshold. This means that even an extremely weakly contagious pathogen could trigger a massive infection wave. However, neither are real-world networks infinite nor are their degree distributions true power laws, so that such predictions cannot be taken at face value but have to be reinterpreted cautiously. The statistician and popular author Nassim N. Taleb has recently analyzed records of historical epidemics from this perspective to find that, indeed, the  \emph{distribution of fatalities is strongly fat-tailed},  ``suggesting a tail risk that is unfortunately largely ignored in common epidemiological models''~\cite{TailRiskDiseases}. 

This goes against most people's common sense, derived from experience with near-equilibrium processes, to which the law of large numbers and the central limit theorem apply. But their intuition is a poor guide far from equilibrium. In particularly, for  (locally) exponential growth processes and processes with considerable tail risks, the coarse-grained variables used to formulate rate-equation models may themselves become scale-dependent, fuzzy concepts, if not mechanistically meaningless. This applies, in particular, to the standard estimators of epidemiological or financial risks derived from these variables, such as the average reproduction number $R$~\cite{InfectionDynamicsScaleFreeNetworks, PropertiesHighlyClusteredNetworks} or the infamous ``value at risk'' (VAR)~\cite{Rebonato}.  For example, rather than controlling the global increase of infections in an epidemic, as assumed by the standard models, $R$ acquires a vague meaning, such as a ``probability for certain large but localized outbreaks'', in heterogeneous networks. And the naive trust of financial-risk regulators in VAR and other average numbers has even been denounced as a key factor in the financial meltdown they were expected to mitigate! Contrary to formal predictions by idealized models, originally devised as ``illuminating caricatures''~\cite{anderson-may:1991} to broadly outline some general principles, local details (like those triggering a traffic jam out of the blue), finite-size effects (how big it gets), and pure chance (when or where it happens) will thus matter a lot, under practical circumstances. Certain real-world epidemics may then quickly go extinct, while others blow up without a clear logical explanation --- no matter how much the fallible human mind clings to reassuring and self-complacent narratives of simple (linear) cause and effect. 

The Petersburg gamble may again be invoked to illustrate the main point behind this perplexing risk management paradox --- and our innate incompetence to appropriately deal with small fatal risks. It explicitly shows how common intuitive measures employed to quantify and manage the course of epidemics or financial markets, such as averages and variances, may become uninformative and unreliable, if not outright deceptive, in the presence of power-law risks. You could say that, as always in statistics, knowing that the average chance for a disaster is low will do little to console you if you are the one who happens to be hurt. But, as the example illustrates, for systems with substantial tail risk, this effect is grotesquely exaggerated far beyond the scale of individual cases (recall that the fair price and average payoff are infinite, but how incredibly rich the banker may get by charging only a finite price). For epidemic disease spreading, this tendency is aggravated by the propensity to produce locally exponential growth. It entails a sensitive dependence on initial conditions and rate coefficients, familiar from chaotic systems, where it is often referred to as \emph{the butterfly effect}.  Its consequence in conjunction with the outlined extreme  heterogeneity effects is the \emph{de facto breakdown of determinism on global length and time scales}. The bottom line is, that the spread of real-world epidemics is far more sensitive to finite-size effects and accidental details, down to the scale of individuals~\cite{WhenIndividualBehaviorMatters} if not within any single organism~\cite{2PercentCarry90Percent,singanayagam:2022,graudenzi-etal:2021,tonkin_hill-etal:2021}, than admitted by the widely adopted theoretical models. Any two outbreaks, even in what appears to be the same setting, may thus have very different epidemiological outcomes~\cite{NetworkTheoryAndSARS,TailRiskDiseases,viboud-etal:2016}.  How far these heterogeneities can judiciously be averaged out~\cite{feng-etal:2019,diekmann:2000,zonta-etal:2021} and subsumed into effective parametrizations~\cite{rudiger-etal:2021,husemann-etal:2016}, to be dealt with deterministically on large enough scales, thus allowing for  ``typically'' (yet not necessarily always) reliable predictions, is an intriguing open question.

\section{Disease control paradox}
As an immediate practical consequence, a highly erratic and somewhat counter-intuitive response to  mitigation measures can emerge~\cite{NetworkTheoryAndSARS,DiscontinuousEpidemicTransitionLimitedTesting}. In particular, while processes with a  narrow-tailed distribution respond favorably to measures acting on bulk behavior (such as a speed limit for all drivers), those with fat-tailed risks respond poorly~\cite{rudiger-etal:2021} and instead \emph{require more specifically targeted interventions}, which, in return, will incur \emph{substantially lower economic costs}. For example,  convincing sex workers to use condoms is vastly more effective in reducing the spread of sexually transmitted diseases than announcing an overall sex ban.\footnote{To the referee who swears by the latter: for how many generations should it be maintained?}
Analogous conclusions apply to other diseases with fat-tailed offspring distributions, such as COVID-19~\cite{EvidenceCoronavirusSuperspreadingFatTailed} or Ebola \cite{EbolaVirusHealthCareWorkers}, which was found to be contracted by health-care workers in Sierra Leone with a 100 times increased risk. Ironically, in epidemics, many of the networks on which our modern societies so heavily rely, may fall prey to their very strength. It is exactly their strongly heterogeneous contact distributions,  leading to transmission graphs as the one shown in the right panel of Fig.~\ref{fig:inftree}, that are responsible for the much-praised resilience of scale-free networks against failure. (Deleting a node is inconsequential unless you have very accidentally hit a hub.)  But in the context of epidemiology, this translates into a vexing  insensitivity of the spreading process against all types of  pharmaceutical and non-pharmaceutical ``bulk'' mitigation measures that are indiscriminately applied to a population as a whole rather than to targeted individuals, specific locations, or critical events~\cite{rudiger-etal:2021,chang-etal:2021}. Which in turn helps to explain why the effectiveness of such measures is so hard to accurately evaluate~\cite{chin-etal:2021}. And why it regularly turns out disappointing in terms of the cost-benefit ratio~\cite{LockdownReportAllen},  even under  highly controlled laboratory settings~\cite{TransmissionAmongMarineRecQuarantine}.   That  each generation of epidemiologists apparently needs to learn this lesson anew~\cite{BiosecurityDiseaseMitigationMeasures,herby-etal:2022}, underscores the immensely unintuitive character of the heterogeneous epidemic dynamics.  

As in the Petersburg paradox, one cannot assume heterogeneities to average out smoothly in epidemiology, as tacitly  done when one writes down coupled rate equations for average variables, claiming simple causal relations between them. Even if such models may nicely capture certain aspects of the observed dynamics \emph{on a descriptive level}, this should not be mistaken as proof of a \emph{causal relation}. To use a hackneyed analogy: a correlation between the numbers of  babies born in a house and storks on its roofs may have a causal origin (e.g., a healthy rural ecosystem may be favorable for both), but this does not imply that more births will mechanistically attract more storks, or \emph{vice versa}. If stated so bluntly, this may seem an obvious pitfall that is easily avoided. But a cursory survey of the literature suggests it to be a rather common trap. Encouraged by the ease with which one can come up with phenomenological fits to the stereotype patterns of infection waves, many authors have apparently failed to realize that the mathematically toxic mix of heterogeneities with locally exponential growth dynamics may render forecasting and forecasting-based risk management and disease control rather subtle tasks, even retrospectively!\footnote{For introductions see e.g.\ T. Wiethölter, 2022: \url{http://coronakriseblog.wordpress.com}; C. Kuhbandner, Telepolis 2020: \url{http://www.heise.de/tp/features/Warum-die-Wirksamkeit-des-Lockdowns-wissenschaftlich-nicht-bewiesen-ist-4992909.html} (German)} Associated common fallacies of data analysis are widely documented and supported by empirical surveys, which underscores the ``need to shift study design toward prioritizing the handling of data sources rather than refining models''~\cite{ForecastingFailed,chin-etal:2021,kuhbandner-etal:2022, AssessingMandatoryStayAtHomeEffectsOnSpread,atkeson-etal:2020}. 

Physicists, as well as meteorologists working on weather forecasts~\cite{MessierThanWeather}, should actually not need to be reminded of this. Weather forecasts can at least rely on real-time data gathered from thousands of automated weather stations all over the land, sea, and sky to rapidly feed back into the models, as well as on a broad consensus on the underlying physical equations. 
And, importantly, the public reactions they trigger (e.g., leaving the house with or without an umbrella) do not react back onto the weather itself to cause what epidemiologists call the ``prediction paradox''. From this distinction, an increased responsibility of the epidemic relative to the meteorological modeller ensues, as pointed out in Ref.~\cite{kuhbandner-etal:2022}.  Based on an in-depth study of a real-world example, it exemplifies how self-fulfilling prophecies can be a particularly harmful type of predictions.

In summary, since deterministic modeling and forecasting is a mine field, responsible risk assessment in epidemiology  hinges most crucially on \emph{precise and unambiguous empirical data}, which are then readily condensed into easy-to-use empirical risk calculators for medical doctors and the general public.\footnote{See, e.g., the QCovid risk calculator: \url{http://qcovid.org}; its current risk estimate for catching COVID and dying from it is 0.003\%  for the author and 0.0001\% for his kids.}

\section{Genomic superspreading}
While a pathogen is invading a host population, mutations,  insertions, and deletions are accumulating in its genome, which gives rise to genetic drift (the technical term for diffusion in gene space). The ensuing coupled spreading dynamics has been a central topic in population genetics and immunology for almost a century~\cite{luria-delbruck:1943}.   Genome sequencing allows to harness it for a new form of spatiotemporal genomic epidemiology.  In this context, genetic changes play the role of a \emph{molecular barcode} and a \emph{molecular clock}. Pathogen spreading can thereby be  backtracked and displayed in the form of phylogenetic trees. This provides modern epidemiology with entirely new tools,  equivalent to C-14 dating in archeology and fingerprints in forensic analysis.  The molecular clock rate decides what time scales can be resolved~\cite{PopulationGenAndEvEpOfRNAViruses}. In the case of highly conserved DNA viruses, such as  the smallpox virus variola, the approach is suitable to study historic time scales~\cite{duggan-etal:2016}.  Intriguingly, the phylogenetic analysis reveals that the little diversification of major variola lineages that has occurred at all, can be traced back to the 18th and 19th centuries, concomitant with the development of modern vaccination. At the other end of the spectrum, if genetic change is many orders of magnitude faster and, as in the case of various RNA viruses, say, occurs on a weekly to monthly basis, with error rates per nucleotide and replication cycle up to the \textperthousand-range, phylogenetic analysis can resolve much finer epidemiological details. It can then help to track the formation of genetic variants leading to the emergence of repeated infection waves~\cite{GenomicReconstructionEngland} and to validate global spreading pathways and local superspreading events~\cite{gomez_carballa-etal:2021} (which show up in the phylogeny in a similar way as in the infection trees in Fig.~\ref{fig:inftree}). It has lately even provided hot trails for contact tracing~\cite{GenomicEpidSuperspreadingAustria}.

Phylogenetic analysis has furthermore backed up estimates for  transmission properties such as the  bottleneck size (the number of virions required for an infection), supporting the notion of substantial transmission heterogeneity, as even a few virions appear to suffice for a successful infection~\cite{GenomicEpidSuperspreadingAustria}. Another interesting  recent finding was the low genetic variability within superspreading clusters. During the early stage of spreading of COVID-19 in the United States, a superspreading study analyzing 118 infection cases in a skilled nursing facility essentially found a monoclonal genealogical fingerprint (59 of the genomes were found to be identical) and a somewhat larger, though still relatively low, genetic variability in a larger superspreading event at a  conference~\cite{BostonSuperspreading}.  These results underscore that superspreading events may exhibit rather little (if any) immune escape. It is indeed plausible that new genetic variants are not bred through superspreading, if it is understood to mean that many individuals are infected by a single person. Genetic variation is more likely to increase by sequential passage through many immunologically diverse individuals. (This is exactly how gain-of-function experiments are conducted, after all~\cite{lipsitch:2018,ArisenViaSerialPassage}, which makes one wonder whether it is really always a good idea to follow the intuitive impulse to ``flatten the curve''.\footnote{\label{ftn:ice}or whether ``as many people as possible need to be infected with the virus'', Reuters 2022: \url{http://www.reuters.com/business/healthcare-pharmaceuticals/iceland-lift-all-covid-19-restrictions-friday-media-reports-2022-02-23/}}) After a series of infections, subsequent superspreading events may however play a key role in selecting and amplifying some of rare genetic variants, and in distributing them widely through the population, while others erratically die out~\cite{VariantEvolutionFounderEffectsBursts,gomez_carballa-etal:2021}, similarly as discussed above for whole epidemics. 

In summary, much as the spreading of pathogens in real space, also their \emph{genetic dynamics is highly nonlinear and governed by substantial heterogeneity}.  It is dominated by the combined effects of repeated passages through narrow population bottlenecks that can produce \emph{mutational bursts}, followed by \emph{proliferation bursts} within the infected organisms and \emph{superspreading} among them. The spread and amplification of emerging variants effected by this repeated chain of events goes under the name of ``founder effect'' in genetics~\cite{VariantEvolutionFounderEffectsBursts}.  As an important consequence of the population-genetic heterogeneity, the mitigation of epidemics of fast mutating pathogens (such as RNA viruses) by pharmaceuticals or vaccination is a very subtle task. Modeling suggests that a widespread distribution of vaccines that afford imperfect immunization and sterilization during an ongoing epidemic may result in a wide range of potential epidemiological and evolutionary outcomes~\cite{EpidemiologicalConsiderationsVaccineDosingRegimes}. In particular, attempts to suppress pathogen spreading in a population by vaccination can promote spreading in gene space. Similar as with antibiotic resistances, infections after waning immunity would be key potential contributors to viral immune escapes via so-called antigenic drift, which may possibly already set in at relatively low vaccination rates~\cite{chong-ikematsu:2018}. This is currently also a  concern in the context of COVID-19~\cite{harvey-etal:2021,prevost-finzi:2021,liu-etal:2021}.  It is therefore imperative to determine the strength and duration of clinical protection and sterile immunity in order to avoid negative impacts. And, similarly as with other mitigation strategies, it is crucial to address the population heterogeneity and selectively target pertinent cohorts to avoid compromising the potential benefits, e.g., by incubating escape mutants in non-vulnerable cohorts~\cite{PreactivatedAntiviralInnateImmunityChildren,borch-etal:2022}.

\section{Survival of the flattest and the flu}
While pathogen superspreading is commonly ascribed to individual supershedders or socially highly connected hubs, there appears to exist another, more subtle and more  worrisome type of superspreading events for rapidly mutating germs. It is characterized by close encounters of many infectious and susceptible individuals under incubator-like conditions, say, at overcrowded parties or similar ``3C''$^{\ref{ftn:3C}}$ events. In terms of network theory, one thus deals with (temporarily) strongly interconnected host clusters, which favor co-infections~\cite{tonkin_hill-etal:2021}, so that one may also speak of  ``cluster infections''. Under such conditions, which do not rely on individual superspreaders~\cite{SuperspreadingWithoutSuperspreaders,gomez_carballa-etal:2021}, both the overall pathogen concentration and its genetic diversity may peak simultaneously, thus likely resulting in burst-like outbreaks both on the population  \emph{and} genetic level. This has important epidemiological and genetic consequences. Namely, on the one hand, it means that a state that would be classified as epidemiologically stable  by the conventional test-tube models (that replace heterogeneous distributions by averages) gets a great chance to conspicuously reveal its actual metastability$^{\ref{ftn:lucky}}.$  On the other hand, it implies that the larger genetic diversity of the invading pathogen ensemble provides it with exceptional opportunities to cope with the immunological diversity of the host population and, moreover, with a unique platform for  \emph{genetic competition and cooperation}. There is then a greatly increased chance for synergistic mutations, reassortments, and recombinations to bring together independent mutations, akin to what happens inside immunocompromised patients~\cite{ParallelEvolutionInfluenzaMultipleSpatiotemporal}. Pathogen evolution is thereby considerably accelerated, and viable new variants may emerge via so-called \emph{antigenic shift}~\cite{bolton-etal:2021}, which would otherwise be quite unlikely.  

To distinguish such incubator-type infections from superspreader transmissions involving supershedders and hubs, alone, one could characterize them as  \emph{cooperative superspreading} (or ``superbreeding'') events~\cite{vignuzzi-etal:2006}. Instead of a monoclonal pathogen meeting a single host organism, such superbreeding involves the encounter of a genetically diverse pathogen population with  a similarly diverse population of immune systems. It is an intriguing speculation, whether the associated  population-genetic  heterogeneities and nonlinearities could possibly give rise to so-called \emph{explosive growth} regimes (e.g., during the carnival week), similar to what is predicted for certain contagious social spreading processes~\cite{iacopini-etal:2019}.  Could this then possibly also help to explain a sudden burst-like emergence of pathogen subtypes and variants that can largely evade the pre-existing and vaccine-induced immunity in their host population~\cite{VariantEvolutionFounderEffectsBursts,vignuzzi-etal:2006,liu-etal:2021}?


On a more conceptual note, in such situations, where (back-and-forth) mutations are frequent, the fitness of an individual genotype becomes meaningless and the target of natural selection is no longer the fastest growing replicator but rather a whole cloud of mutants (of continuously changing prevalence). Of which the catalogued subtypes and variants represent merely the tip of the iceberg~\cite{domingo-perales:2019}. This cloud is the so-called quasispecies, which can collectively outperform single competing genotypes, even fitter ones, via the ``\emph{quasispecies effect}'' or ``survival of the flattest''~\cite{DigitalEvolutionSurvivalFlattest}. Put simply, having too many mutations in a genome, most of which will be deleterious or at best neutral, can drive a population to extinction (``Muller's ratchet''). But also too few mutations can cause extinction, namely by rendering a population unable to survive changes in the environment, and to recover its genetic diversity after drastic population reductions in bottlenecks~\cite{SmerlakBottleneck}.  Quasispecies evolution thus allows fast mutating microbes to strike an optimum balance between evolvability and structural preservation~\cite{domingo-perales:2019}. Importantly, the quasispecies clouds should not be imagined as simple collections of mutants but rather as groups of interactive variants, which reappear after each bottleneck, via a stupendous replication burst that may boost their numbers from a handful of virions to many billions. This allows for an effective sampling of large parts of the available sequence space, and concomitant phenotypic expansions, so that each quasispecies genome swarm in an infected individual is unique and new~\cite{schneider-roossinck:2001}. Each genetic and phenotypic pattern then contributes collectively to the characteristics of the microbial population, helping its resilience and colonization of complex ecosystems~\cite{vignuzzi-etal:2006}. 

Many consequences of quasispecies dynamics run counter to traditional views of microbial behavior and evolution and have profound implications for the understanding of viral disease. An example is the apparent paradox that attenuated RNA viruses from vaccines can occasionally revert to more virulent forms, thereby revealing their ``quasispecies memory''.  A possible case is polio~\cite{vignuzzi-etal:2006}, for which the World Health Organization has over the last few years consistently reported more vaccine-derived than wild-type cases. Another example of quasispecies dynamics at work should be the above mentioned cooperative superspreading events~\cite{VariantEvolutionFounderEffectsBursts}. They may be suspected to play a major role in the spread of various viral diseases~\cite{SuperspreadingWithoutSuperspreaders,gomez_carballa-etal:2021,domingo-perales:2019}, including the seasonal flu, which infects between 5 and 15\% of the world’s human population every year, causing several million cases of severe illness, and hundred thousands of respiratory deaths. As previously for SARS and MERS, sequencing  indicates that the quasispecies concept also applies to SARS-CoV-2.  A considerable intra-host genetic heterogeneity was detected~\cite{graudenzi-etal:2021}, and quasispecies were seen to differ from one day to the next, as well as between anatomical sites~\cite{ViralQuasispeciesDuringSarsCov2}. \emph{In vivo}, these viruses thus all appear as complex and dynamic variant distributions (not properly resolved by oral or nasal swabs alone). So they are potentially capable of producing staggering cooperative mutational bursts. It remains an open question whether these can account for the mystifying assembly of  omicron~\cite{liu-etal:2021}. 

\begin{figure}
    \centering
    \includegraphics[width=\textwidth]{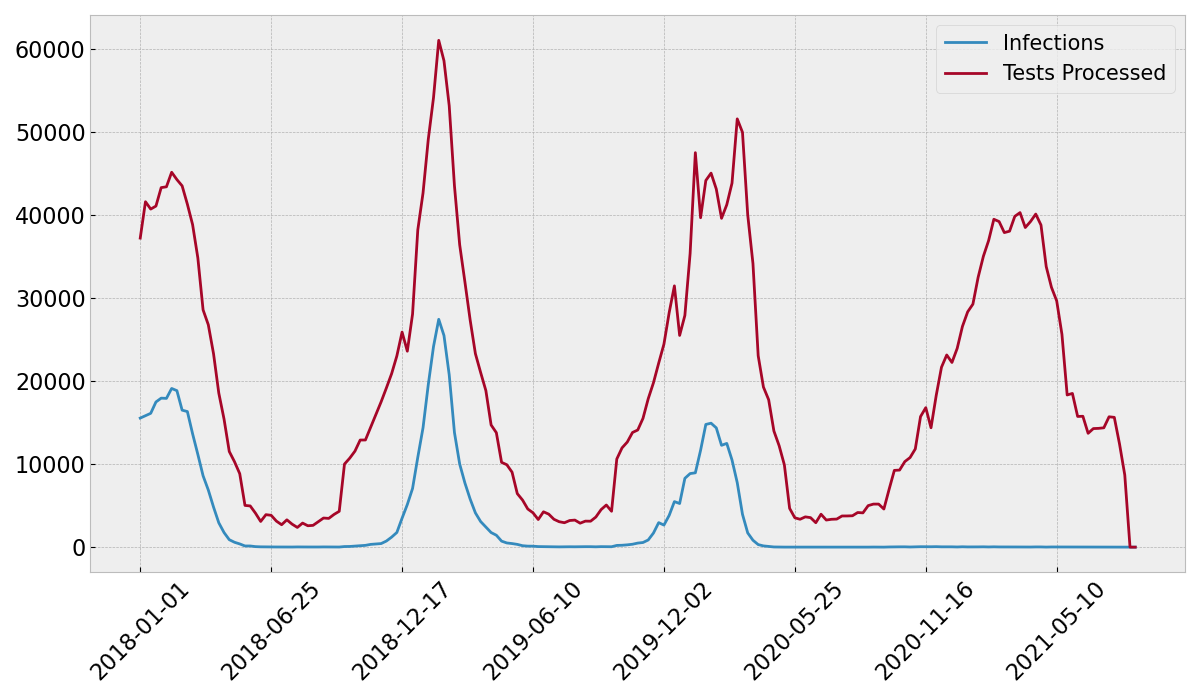}
    \caption{\textbf{Flu tests and positive cases} (here for the European WHO region$^{\ref{ftn:flu}}$) indicate that the flu has just been eradicated --- or has it? Quasispecies and other models of eco-evolutionary dynamics may suggest otherwise.}
    \label{fig:FluTestEurope}
\end{figure}

It should not come as a surprise that the development of effective sterilising vaccines to halt quasispecies epidemics is notoriously difficult, as testified by empirical studies under  well-controlled settings for both human populations~\cite{USNavyInfluOut} and their livestock~\cite{HerdImmunPoultry}. Even if vaccination reduces the number of influenza infections by a particular strand, meta-analysis cannot detect a systematic reduction of the overall number of influenza-like illness episodes~\cite{jefferson:2006,VaccinesForHealthyAdults}. Other pathogens or variants seem to happily fill the gap~\cite{InfluenzaLikeIllnessNotRed, IncreasedRiskNoninfluenzaInfection}. 
Enthusiastic claims that ``real-time practical forecasts'' of the seasonal flu can be achieved ``by leveraging historical and modern experimental assays and gene sequences''~\cite{huddleston-etal:2020} to improve vaccination and disease control, have to be gauged against this bleak empirical evidence, as well as against Fig.~\ref{fig:FluTestEurope}\footnote{\label{ftn:flu}World Health Org. 25.08.2021: \url{http://apps.who.int/flumart/Default?ReportNo=12}}. It exemplifies (for Europe) that the worldwide seasonal flu has seemingly just been extinguished. The effect appears to be insensitive to a country's ``stringency index'' or ``containment and health index'', and neither vaccination nor forecasting can claim credit for it. 

One can conclude that, with fast mutating RNA viruses, vaccinologists face similar heterogeneities and nonlinear effects (yielding unexpected and counter-intuitive outcomes), on the level of a virus population, as epidemiologists experience with measures targeting cohorts of its host population~\cite{suzuki:2006}. They are recently particularly worried by a possible negative impact of repeated influenza vaccination on vaccine responsiveness and vaccine effectiveness (protection from infection)~\cite{khuruna-etal:2019}. And they warn that with improved vaccine efficacy and wider application comes a heightened risk for immune escape, thus likely creating the right conditions for a significant zoonotic transfer leading to a pandemic~\cite{bolton-etal:2021}.  To reduce such risks, current flu-vaccine developments aim at highly conserved parts of viruses, and, on the same grounds, researchers advise against antiviral monotherapy (use of a single antiviral agent)~\cite{domingo-perales:2019,kundu-etal:2022}. Other major examples of the stunning ability of biological systems to defeat our attempts at mitigation in similar ways include the worldwide spread of antibiotic resistance genes across distantly related bacteria, crossing species, and phylum boundaries and physical locations; the rapid evolution of cancers in the face of chemical attack; and the ability of HIV to out-adapt treatment. Which prompted the authors of Ref.~\cite{LifeIsPhysics} to draw a bold analogy between the fundamental limitations to current medical practice and attempts to design integrated circuits without a fundamental knowledge of quantum electronics and semiconductor physics.

\section{The eco-evolutionary perspective}
The foregoing paragraphs have provided a glimpse as to why there has recently been increasing mutual interest and synergies between the fields of epidemiology and eco-evolutionary dynamics. 
Thanks to the rapid recent development of micro-manipulation techniques, sequencing, and synthetic biology, epidemiologists can now increasingly resolve and harness the crucial dynamics in the ``genomic sector'' of the complex, multi-layered process of disease spreading. In return,  the traditional, descriptive approach to evolution has increasingly given way to a bona-fide experimental science. The modern perspective on evolution has thereby naturally become more microbe-centric, and better aligned with the fact that most of life is microbial, and all life primarily relies on microbial processes. And it has embraced the notion that ecological and evolutionary time scales may often mutually interfere much more than admitted by some core postulates of the traditional theories of evolution (and epidemiology, for that matter) in terms of the ``Central Dogma'' or ``Modern Synthesis''~\cite{IllusionModernSynthesis}. Microbial species have a number of unique characteristics that are not captured by conventional ecological and evolutionary theory, chiefly their rapid evolution, horizontal gene transfer, their ability to produce public goods, toxins and antibiotics,  and their complex feedback with hosts~\cite{MicrobiomeBeyondHorizon}.  This complexity may partly be attributed to genetic variability, but recent research has moreover underscored a remarkable ability for phenotypic diversification~\cite{schroeter-dersch:2019}.  

Progress in the field is now boosted by real-time laboratory studies of the eco-evolutionary dynamics of bacteria and viruses, which are ideal models for that purpose. Physicists (in particular statistical mechanicians) have become major contributors, eager to exploit interdisciplinary synergies between the converging fields of mathematical and evolutionary biology, ecology, epidemiology, game-theory, and non-equilibrium statistical physics~\cite{LifeIsPhysics}. Their ambitious long-term aim is to create a fundamental understanding of how the unique self-referential dynamics of evolution (in which the conventional distinction between the physical data and its governing laws is largely abolished) arises as a universality class from the molecular processes, via an extreme version of non-equilibrium statistical physics. This development nurtures hopes for future progress in epidemiology and vaccinology that could lead to a much deeper level of understanding and possibly to  smarter, scientifically better grounded mitigation strategies.

\section{Lessons from frustrating games}
Apart from being confounded with gene spreading,  epidemic spreading is also closely intertwined with information spreading. As more extensively discussed in the contribution by Platteau and co-workers in Chapter~22 of this volume, game theory can help to address many of the difficult, multi-layered problems resulting from this additional complication. Here, I want to focus on a particularly interesting aspect, namely information feedback with frustration.

\begin{figure}
    \centering
    \includegraphics[width=\linewidth]{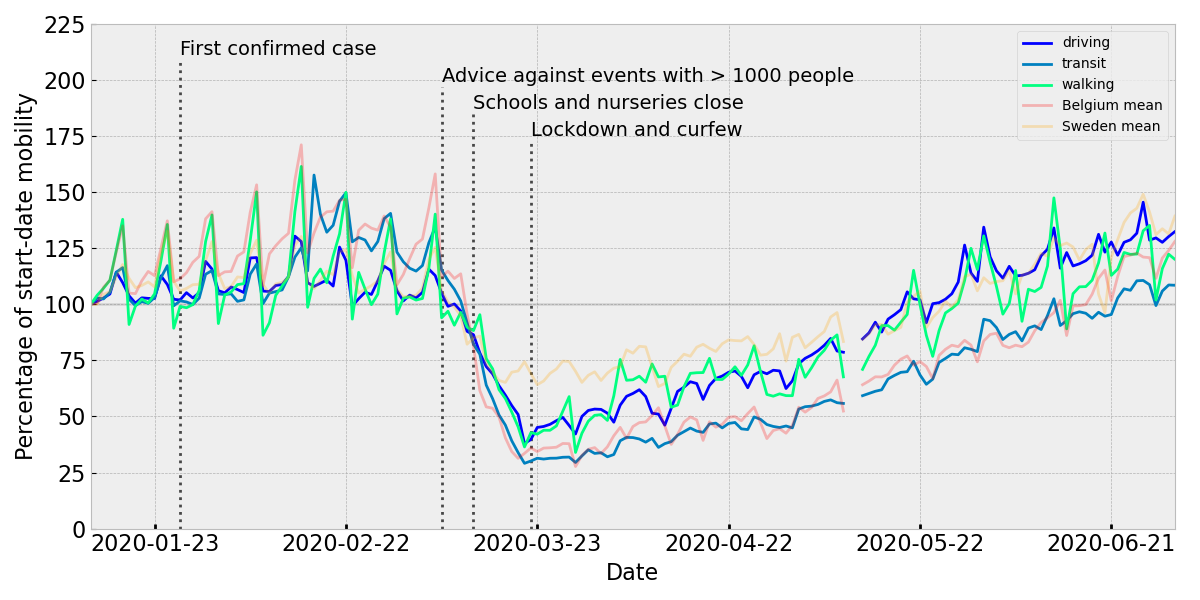}
    \caption{\textbf{Routing requests during the early COVID-19 epidemic} (Apple Maps, normalized to starting date) are  proxies for human mobility. Different means of transport and major government interventions (dotted lines) for Germany are compared to aggregated data for Belgium and Sweden. Throughout Europe, \emph{news coverage} of the emerging pandemic soared drastically$^{\ref{ftn:media}}$  from March 10 to March 20, in perfect correlation with the  decreasing mobility.}
    \label{fig:MobDat}
\end{figure}

First, to briefly clarify the relevance of \emph{information feedback}, let me note the obvious: while pathogens spread through a population, its members may acquire and evaluate knowledge about the spreading process and adjust their behavior accordingly~\cite{DynamicalInterplayAwarenessEpidemicSpreadingNetworks}.  That animals can discriminate infectious individuals by their scent~\cite{ScentDogIdentification} suggests that this feedback effect is not limited to human populations.   It allows individuals to act according to their own judicious risk assessment. Such information flow between agents and/or feedback about the state of the community can generate very complex dynamics, where small causes can have disproportionately large effects~\cite{StabilityDiversityCollectiveAdaptation}.  Mathematically speaking, one deals with a layered process of (piggyback) disease spreading plus superimposed information spreading. The latter acts very similarly to a competing infectious strain in a  multilayered (``multiplex'') network, as discussed in Chapter 19 of this volume.  And there is indeed new evidence from mobility surveys to support the notion that it is actually information  about the spreading process (rather than centrally imposed mitigation orders, say), which arguably has the strongest impact on pertinent behavioral change during an epidemic --- see, e.g., Fig.~\ref{fig:MobDat}\footnote{\label{ftn:media}F. Bol{\'i}var \emph{et al.},  BBVA Research 2020: \url{http://www.bbvaresearch.com/en/publicaciones/monitoring-covid-19-pandemic-using-big-data-from-the-media}} and Refs.~\cite{ImpactOfSarsCov2MobilityGermany,LongDistanceTrains,chang-etal:2021,watanabe-tomoyoshi:2021}.
In this respect, the infected population may be viewed as an agent-based algorithm, employing swarm-intelligence in order to collectively outsmart the spreading pathogen, in a self-organized way~\cite{MoreDataToIndividualCitizens,watanabe-tomoyoshi:2021}. This distinctive feature of epidemiological dynamics has various profound practical, social, and political implications, further elaborated in Sec.~\ref{sec:infodemics}.  

 \emph{Frustration} is a special characteristic feature deeply ingrained in epidemiological information feedback, though maybe not quite so obviously. Game theoretic tournaments have proven  follow-the-crowd or herding strategies of social copying to be surprisingly successful~\cite{WhyCopyOthers}. In particular, they may drive swarms or flocks towards a so-called non-equilibrium critical state, with a large susceptibility to external stimuli. The latter may help swarms to defy predators by generating large-scale coherent response patterns~\cite{CriticalityDynScalingLivingSystems}. However, in the case of disease spreading, such herding, which ``aligns'' individuals, usually  with some time delay~\cite{holubec-etal:2021}, is not at all a smart reaction. In this context, swarm formation  essentially either describes the epidemic spreading of the pathogen  itself, or  some equally unfavorable panic reaction in the host population. Indeed, epidemic diseases are known to trigger psychological and sociological patterns in human societies that can themselves be analyzed in epidemic terms~\cite{strong:1990,aiello-etal:2021}. {\interfootnotelinepenalty10000 Helped by the dominant fear culture in the media~\cite{chaiuk-etal:2020}\footnote{For Germany, see e.g.\ D. Gräf, M. Hennig: \url{http://www.researchgate.net/publication/343736403_Die_Verengung_der_Welt_Zur_medialen_Konstruktion_Deutschlands_unter_Covid-19_anhand_der_Formate_ARD_Extra_-Die_Coronalage_und_ZDF_Spezial}}}, such psychic epidemics can seize whole societies, undermine and disrupt their social fabric, and make them slip into primitive collectivism and so-called \emph{mass formation}~\cite{esfeld:2021,desmet:2022}. 

How would a smart reaction instead look like? From the perspective of an infected population, quantifying the ``best'' counter-strategy and the ``cost'' of an epidemic can become exceedingly difficult, due to the  high dimensionality of the notion of cost and its heterogeneity across the population. As discussed above, the specific threats due to the spreading of the disease may be very unevenly distributed throughout a population. Furthermore, they need to be gauged in a context of other diseases~\cite{bell-hansen:2021} and the economic and social costs of mitigation measures~\cite{BiosecurityDiseaseMitigationMeasures,LockdownReportAllen}, which may be distributed quite differently but also very unevenly. Therefore, it is highly unlikely that a globally applicable one-fits all optimum solution exists. This introduces the crucial element of frustration, as one would say in physics, suggesting that the host population might try out a reciprocal \emph{strategy of decoherence}. Frustrated systems stand out in that the desired optimum cannot be reached by any uniform rule but instead requires a broad diversity of strategies. In this respect, they are reminiscent of glasses and spin glasses in physics, which have challenged condensed-matter physicists for decades and still remain enigmatic (despite all theoretical progress honored by the 2021 Nobel prize in physics\footnote{For which the Nobel prize committee justly emphasizes ``the vastness of the landscape of disorder'', \url{http://www.nobelprize.org/uploads/2021/10/sciback_fy_en_21.pdf}}). Briefly, if an ensemble of magnetic moments suddenly and spontaneously undergoes a phase transition to a state of collective alignment upon cooling, one speaks of a ferromagnet.  This is the physical analogue to the mentioned groupthink strategy of social copying, and the outcome is mass formation~\cite{desmet:2022}.  However, in strongly disordered magnets (as in real-world epidemics), global alignment is not a favorable option. One then speaks of a frustrated system. The magnetic moments start to search and compete for very complex, disordered low-energy configurations in a highly degenerate, rugged free-energy landscape, which entices them into exceptionally slow, non-stationary dynamics. And this description comes much closer to the problem faced by a population invaded by a pathogen than a ferromagnet.

To break down the essence of frustration to a comparatively simple example, consider the so-called \emph{minority game}~\cite{challet:2005}, 
 which was originally devised to elucidate the workings of market economies. It can, with some qualification, indeed formally be mapped onto a spin-glass problem and has been studied analytically by the replica method \cite{martino-marsili:2001}. In this game, each participating agent has only two options to choose from, which may naturally be named ``go'' (i.e., go out to get work done, food organised, etc.) and ``stay'' (i.e., stay at home to avoid overly crowded situations), in an epidemiological context. The spreading pathogen would clearly ``prefer'' its host agent to join the majority or best follow a globally coherent rule, in order to optimize its spreading. The agent, instead, tries to anticipate the future moves of everybody else in the game, in order to decide how to best end up among the minority (going out when streets are empty, staying home when they are crowded). 
To this end, each agent is endowed with a set of (random) strategies or predictors that guide its decision to stay or go. A strategy rises or falls in esteem, based on whether it gets the individual into the minority or not. In this way, each agent builds a personal  decision matrix, which may evolve over time, according to its experience. 

Intriguingly, if the game is played on a computer or with human players, it turns out that the population soon self-organizes so that attendance fluctuates near an optimum value, without external guiding. The fact that the agents use different strategies to predict what the majority will do, is however essential for the spontaneous emergence of a smart solution. There is no ``representative agent'', and what matters is the interaction between all the agents’ individual decision making processes. \emph{The best  solution arises from the interplay of many diverse individual strategies}. In fact, the quality of the solution deteriorates if too many agents align their strategies, which may give rise to so-called pork cycles (everybody staying home today because the streets where crowded yesterday). Even worse, if the population gets dominated by trend-followers rather than minority seekers, the subtle optimal mode of decoherent cooperativity  may catastrophically collapse \cite{kozlowski-marsili:2003}. Besides, to a lesser degree, it also deteriorates if the predictors become more sophisticated than required by the actual behavioral complexity  of the population as a whole (or, alternatively, if the latter is artificially too much suppressed). In economic terms, the latter limit corresponds to the paradigm of a perfectly efficient market, where smart strategies cannot beat random strategies, and there is nothing to gain by adaptive behavior. 

Translated into the epidemiological context, the take-home message from the minority game is thus very much what was already emphasized above:  for a  population to cope well with an epidemic, it is crucial that all  \emph{individuals  have access to accurate and unambiguous empirical data about the course of the epidemic, and that they can make up and freely follow their best individual risk assessments and optimization strategies}~\cite{MoreDataToIndividualCitizens}. A counter argument that is sometimes raised  (in discord with empirical findings, as it seems~\cite{watanabe-tomoyoshi:2021}) against such highly individualized optimization strategies is that they might fail to protect some particularly vulnerable minorities. However, given their low cost compared to certain excessively expensive, intrusive, and potentially harmful alternative strategies that have recently been tried (lockdowns, emergency mass vaccination programs), they leave ample resources for that purpose.

\section{Infodemics}\label{sec:infodemics}
\begin{figure}
    \centering
    \includegraphics[width=\textwidth]{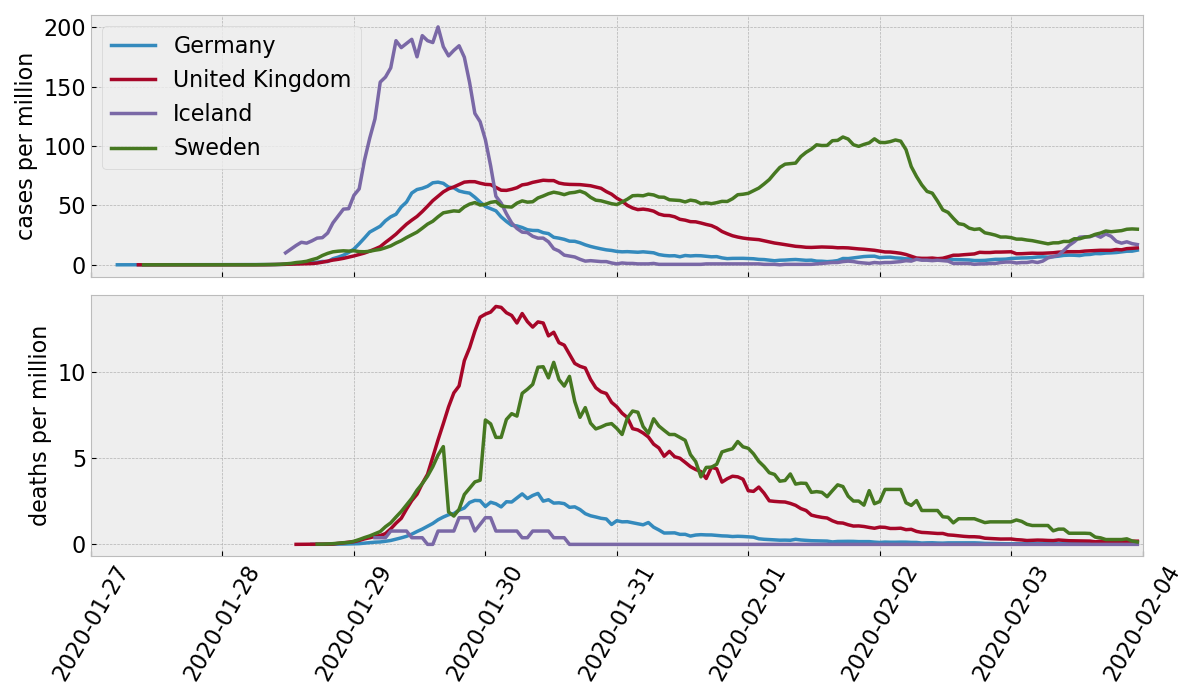}
    \caption{\textbf{PCR positives} as proxies for COVID-19-infections and -deaths per million inhabitants for four European countries (OurWorldInData).}
    \label{fig:CasesDeaths}
\end{figure}

According to the above conclusions, a central task of epidemiologists should be to supply the public with reliable pertinent real-time data about an unfolding epidemic, in the spirit of weather reports. For several reasons, this is not an easy task, however. First, data are often poorly standardized, marred by reporting delays, and scarce and statistically unreliable when most urgently needed, namely during the early stage of a new epidemic. Secondly, the very diverse ways raw data for various parameters are acquired, evaluated, and graphically presented may impede their appropriate interpretation by the public. Figure~\ref{fig:CasesDeaths} may serve to illustrate the point: initially the widely reported case numbers, which were not normalized to test frequency (if to anything at all) suggested that Iceland and somewhat later also Sweden were hit very hard compared to the UK and particularly Germany. Considering instead the cumulative mortality after the first infection wave, one could have concluded instead that Iceland had had no epidemic, while Sweden and the UK had been similarly severely hit. Moreover, by the end of 2021, the daily German death rates turned out to be roughly an order of magnitude above Iceland's and Sweden's, with an all-time case fatality rate (CFR) 30\% above Sweden's. Sweden had taken by far the most relaxed attitude, almost exclusively relying on the voluntary action of its citizens$^{\ref{ftn:weight}}$, a policy eventually adopted by Iceland$^{\ref{ftn:ice}}$ when its so-called 7-day incidence was close to $4\times$  Germany's (Fig.~\ref{fig:incidence}) and $25\times$ Sweden's. So, what should the man in the street make of all this?

Also, whether a country's outcome of an epidemic depends on its government and media leaning more toward the threat-denial or the fear-mongering side, seems a matter of debate~\cite{painter-tian:2021}. For example, the Russian government was accused to have contributed to the country’s comparatively high number of COVID-related excess mortality (possibly as high as 20\% in 2020~\cite{timonin-etal:2022}) by downplaying the COVID threat. But such accusations would probably not be raised against the Spanish government, although Spain's excess mortality (about 14\%~\cite{kowall-standl-etal:2021}) was not so far off. In any case, the communication between governments and the public could be prone to considerable nonlinear effects. As a case study plausibly suggests, unconvincing or inadequately communicated government rigour, however well-intentioned, risks to trigger adverse reactions of their citizens that can exacerbate the impact of an epidemic~\cite{kilmovsky-nemec:2021}. In this light, the lockdowns of a whole country (New Zealand) and parts of one of the world’s largest container ports (China), each \emph{in response to a single positive case} in August 2021, may have been counterproductive, as they conveyed an unrealistically exaggerated threat to the public; especially, if gauged against the low or even insignificant excess mortalities in countries that coped with a range of far less drastic, mostly voluntary interventions, such as Japan~\cite{watanabe-tomoyoshi:2021} and Sweden$^{\ref{ftn:weight}}$ (or also Germany, for which a government-commissioned investigation found that the health system was at no time in danger of being overwhelmed\footnote{B. Augurzky \emph{et al.} 2021, \url{https://www.bundesgesundheitsministerium.de/presse/pressemitteilungen/2021/2-quartal/corona-gutachten-beirat-bmg.html}})~\cite{kowall-standl-etal:2021}.

On top of all these perplexing and seemingly inconsistent observations, there are other, less fateful problems, mainly related to poor data handling and presentation, against which politicians, the media, and even health-care officials\footnote{See, e.g., R. Hughes, UnHerd 2022: \url{http://unherd.com/thepost/nhs-england-deletes-misleading-covid-stats-video}; and, for Germany, M. Barz, W.I.R. 2022: \url{http://odysee.com/@WIR:b/Marcel-Barz-Corona-und-Zahlen:6}} are apparently not immune. To pick an illustrative example, during the early COVID-19 pandemic, the Süddeutsche Zeitung, a widely read German newspaper, regularly reported the time series of the cumulative PCR-positive cases by plotting them in a fixed square-shaped frame. By the necessary rescaling of the daily expanding time axis, the sub-exponential and eventually saturating growth process ended up being  displayed as a progressively steepening curve.  This malpractice  (which would certainly deserve to be included in one of Edward~R.~Tufte's fabulous books on data visualization) only stopped in late April 2020, not too long before the curve would have degenerated into a meaningless step function. And a systematic empirical study of the mainstream-media's coverage of the first year of the pandemic detected a serious lack of contextualization and awareness of scientific uncertainties and debate, plus a significant bias in favor of  politicians of the ruling parties and hard authoritarian interventions --- prompting the authors to raise the question whether fighting a pandemic can be such an obvious priority as to override the standards of diverse and balanced reporting\footnote{M. Maurer, C. Reinemann, S. Kruschinski. Einseitig, unkritisch, regierungsnah? Eine empirische Studie zur Qualität der journalistischen Berichterstattung über die Corona-Pandemie. (Rudolf Augstein Stiftung, 2021)}. As a rule, journalists, politicians, and the wider public tend to underestimate the substantial uncertainties inherent in scientific modeling and debate, and are prone to trust individual favorable model predictions beyond their range of validity~\cite{WrongButUseful}. Moreover,  in a retrospective study of the British BSE crisis, scholars did not attest politicians who claimed to ``follow the science'' much credibility. Instead, they concluded that ``secrecy allowed ministers and senior officials to maintain the public pretence that [\dots] policies were based on secure scientific foundations, while privately acknowledging that regulations were being decided entirely on political rather than scientific grounds''~\cite{PoliticsExpertAdvice}. 

In the more recent COVID-19 context, although the World Health Organization recommends that, to be able to successfully fight an epidemic, policy makers should ``inform, empower and listen to communities''\footnote{\label{ftn:who}WHO Director-General, opening remarks at media briefing on COVID-19,  3.Aug 2020}, a report by Amnesty International\footnote{AI  2021: Silenced and misinformed --- freedom of expression in danger during COVID-19} finds that, ``governments have curtailed freedom of expression instead of encouraging it''.  
Convinced that tightening surveillance and influencing behaviour are ``central to public policy, and government (\emph{sic}) can draw on a potentially powerful new set of tools''\footnote{\label{ftn:mindspace}P. Dolan \emph{et al.}, Mindspace – Influencing behaviour through public policy, 2010: \url{http://www.instituteforgovernment.org.uk/sites/default/files/publications/MINDSPACE.pdf}}, some politicians have actively sought psychological$^{\ref{ftn:mindspace}}$ and technical\footnote{Chaos Computer Club press release, 2021: \url{http://www.ccc.de/en/updates/2021/offener-brief-alle-gegen-noch-mehr-staatstrojaner}}  expert advice how to spy on and to nudge (rather than better inform) the public in a crisis. And a growing research community is now studying the spreading and artificial-intelligence aided~\cite{MisinformationAdoptionOrRejection}  collection, analysis, and blocking of  misinformation diffusing through the world wide web, in order to fight ``the first global infodemic\footnote{which is associated with exponential growth and overabundance of information and can seriously impact on the course of an epidemic, according to the World Health Organization}''~\cite{alam-etal:2021} and to further develop the art of censorship, in general~\cite{DatasetOfStateCensoredTweets}. Advancing internet censorship\footnote{\label{ftn:cdc-data}\dots and stowing away statistical data, because it is prone to misinterpretation; see, e.g., L. Brownlie, The National 2022: \url{http://www.thenational.scot/news/19931745.covid-data-will-not-published-concerns-misrepresented-anti-vaxxers}; A. Mandavilli, The New York Times 2022: \url{http://www.nytimes.com/2022/02/20/health/covid-cdc-data.html}} can be understood as a type of information-layer equivalent of what lockdowns attempt in real space. The declared aim is to mitigate an epidemic indirectly, via central control of the superimposed information spreading. A telling example is provided by two blogs published by  Tomas Pueyo and Aaron Ginn on March 10 and 20 2020, respectively, on the web-platform Medium. Both blogs quickly went viral, acquiring millions of views within days, thereby spreading awareness of the pandemic. Pueyo advocated lockdowns (to restrain an otherwise allegedly$^{\ref{ftn:lobs}}$ unrestricted exponential growth of infections), whereas Ginn's blog questioned their adequacy, warning of hysteria --- and was deleted within 24 hours\footnote{Y. Weiss, Real Clear Politics 2020: \url{http://www.realclearpolitics.com/articles/2020/05/28/how_media_sensationalism_big_tech_bias_extended_lockdowns_143302.html}}.  While it is quite unlikely (if only due the language barrier) that these two blogs have decisively affected the almost perfectly concurrent sharp turning points in the mobility pattern observed in Germany (Fig.~\ref{fig:MobDat}), censors seem to act as if they did. A common argument is that the public needs likewise to be protected from hazardous misinformation as from biological pathogens. However, an alternative solution, more in line with democratic principles\footnote{BigBrotherWatch: \url{http://bigbrotherwatch.org.uk/2021/09/governments-online-safety-bill-poses-greatest-threat-to-free-speech-in-living-memory-say-campaigners}}, could be to offer filters such as News Guard's  browser extension Internet Trust Tool for safe, supervised news consumption to those particularly scared of misinformation, and let the others read, think, and decide for themselves.  

Above all, this raises the question: \emph{who decides what is right or wrong, trustworthy information, and mis-, mal- and disinformation}\footnote{mis-, mal-, and disinformation are now counted among the \emph{terrorism threats} to the US: \url{http://www.dhs.gov/sites/default/files/ntas/alerts/22_0207_ntas-bulletin.pdf}}~\cite{clarke:2021}? Should it be the  administrators of LinkedIn, who temporarily suspended and then reinstated the account of Robert Malone, who is widely respected for his  seminal contributions to mRNA-vaccine development\footnote{A. Sadler, LifeSite 2021: \url{http://www.lifesitenews.com/news/linkedin-reinstates-mrna-inventors-account-after-deleting-it-for-spreading-misinformation}}? Or those of Twitter who recently suspended his account, to the dismay of half a million followers? Or better YouTube's, who censored the epidemiologist Knut Wittkowski$^{\ref{ftn:censor}}$?   Or is it Facebook's certified\footnote{by the Poynter Institute for Media Studies, whose donors include the Charles Koch Institute, the National Endowment for Democracy, the Omidyar Network, Google, and Facebook~\cite{clarke:2021}, and which seems to fiddle even with the Wayback Machine (e.g.\ H. Buyniski, RT 2020: \url{http://www.rt.com/op-ed/505437-internet-archive-censorship-slippery-slope})} third-party fact checkers, who first banned the so-called lab-leak theory, together with millions of other contents (including the accounts of New York University researchers who investigated such practices\footnote{J. Vincent, The Verge 2021: \url{http://www.theverge.com/2021/8/4/22609020/facebook-bans-academic-researchers-ad-transparency-misinformation-nyu-ad-observatory-plug-in}}), but later  changed their minds\footnote{F. Sayers, UnHerd 2021: \url{http://unherd.com/2021/05/how-facebook-censored-the-lab-leak-theory}}~\cite{CovidLabLeakHypothesis}?
What if the editors in chief of one of the world’s top most cited general medical journals have a point, who, in an open letter  to Mark Zuckerberg, complain about a fact check as ``inaccurate, incompetent and irresponsible''~\cite{bmj-zuckerberg:2021}? Can it be that ``Facebook is trying to control how people think under the guise of ‘fact checking'?''~\cite{coombes-etal:2022} and that such fact checking activity has ``enforced falsehoods about the pandemic’s origins, never evaluated the evidence, never apologized, and was never held accountable''\footnote{A. Rindsberg, Tablet Magazine 2021: \url{http://www.tabletmag.com/sections/news/articles/lab-leak-fiasco}}? Maybe not all too surprisingly, two years into the ongoing pandemic, facing a series of scandals with fishy data handling and communication by corporate and official sources, legacy media write about a ``pandemic of ignorance''\footnote{T. Röhn, Welt 2021: \url{http://www.welt.de/235442252}} and reveal that the US Centers of Disease Control and Prevention (CDC) have withheld critical data on the COVID pandemic for up to a year over fear that the information might be misinterpreted and ``because basically, at the end of the day, it’s not yet ready for prime time''$^{\ref{ftn:cdc-data}}$. Again, whether ``mistrust in the authorities'' is indeed ``a complex problem that needs a holistic approach [\dots] of journalists, fact-checkers, policymakers, government entities, social media platforms'' and demands a ``call to arms''~\cite{alam-etal:2021}, or whether it is rather indicative of their failure to properly ``inform, empower and listen to communities''$^{\ref{ftn:who}}$, seems debatable. 

Finally, also the scientific debate itself does not take place in an ivory tower.  Due to a widespread feeling of urgency, the scientific publishing process has often been accelerated during the latest pandemic~\cite{jung-etal:2021,horbach:2021}, and the exploitation of this trend by scientists ``surfing the COVID-19 scientific wave''~\cite{abbas-pittet:2021} has become a concern. Worse, there is circumstantial, formal, and judicial possible evidence that science is becoming increasingly politicized~\cite{rasmussen:2021} as well as manipulated by governments and corporations, and that ``legitimate tools of regulation within science have become weaponized''~\cite{silencing-science:2021,elisha-etal:2021}.\footnote{See also J.~P.~A.~Ioannidis, Tablet Mag.2021: \url{http://www.tabletmag.com/sections/science/articles/pandemic-science}; I. Birrell, UnHerd 2021: \url{http://unherd.com/2021/08/how-china-could-win-the-lab-leak-debate}; and the archive at \url{http://retractionwatch.com}}   According to the assessment of two renowned experts on business ethics and philosophy of science, these are not merely isolated cases, accidental mistakes, and coincidences. The authors rather see a pattern of systematic political abuse of science~\cite{luetge-esfeld:2021}. In this context, the well documented historical malpractices of the tobacco industry may provide an instructive cautionary paradigm. The release of millions of internal documents as a result of the Master Settlement Agreement in 1998 revealed a systematic manipulation of research on health risks associated with smoking; e.g., by launching a public relations campaign about ``junk science'' and ``good epidemiological practices'' and creating fake controversy under the motto ``doubt is our product''~\cite{TobaccoManipulation}, heavily exploiting the penchant of the press for controversy and its often naive notion of balance~\cite{HistoryOfTobaccoIndustry}.  Similarly aggressive or ``disruptive'' covert marketing patterns have more recently been documented for the health industry, e.g., during the (arguably mislabelled) swine flu pandemic and elsewhere~\cite{schonhofer-schulte-sasse:2012,hutson:2009,routledge:2018,doshi-etal:2022}.  Most recently, documents obtained through a freedom-of-information-act request have prompted accusations that leading virologists colluded to mislead the world with regard to the plausible scenarios for the origin of SARS-CoV-2.\footnote{\url{http://odysee.com/@reitschuster:3/220201-wiesendanger-v1:5}}

Altogether, one may eventually feel drawn to the somewhat paradoxical conclusion that we are facing an information crisis while virtually drowning in information. And in the face of the heterogeneous character and the sheer complexity of epidemic spreading, this can make one worry as to how much all our scientific and information technological advances can in the end really help us to outsmart a major epidemic.

\section{Conclusions}
Throughout this chapter, I have emphasized the effects of phenotypical, social, spatial, and genetic heterogeneities on epidemic spreading. It may thus  be read as a warning that the deceptively simple conventional epidemiological models, in which most of this complexity is swept under the rug, should not be taken too literally.  The bottom line is that, in epidemics, the tail often wags the dog. There is a subtle interplay of eco-evolutionary pathogen dynamics with broadly distributed transmissibility, social connectivity,  and spatial mobility patterns. They render epidemics a manifestation of a very complex dynamical system. A breakdown of large-scale determinism, associated with erratic burst dynamics, superspreading events, high tail risks, and a failure of SIR-type forecasting and global disease management practices, was identified as a potential major consequence. Attempts to contain infectious disease spreading  by pharmaceutical and non-pharmaceutical interventions need to face its multi-layered character: heterogeneous spatiotemporal spreading is intertwined with heterogeneous genetic spreading, and attempts to suppress the former may promote the latter, creating potentially worrisome immune escape variants. 

\begin{figure}
        \centering
       \includegraphics[height=0.52\textwidth]{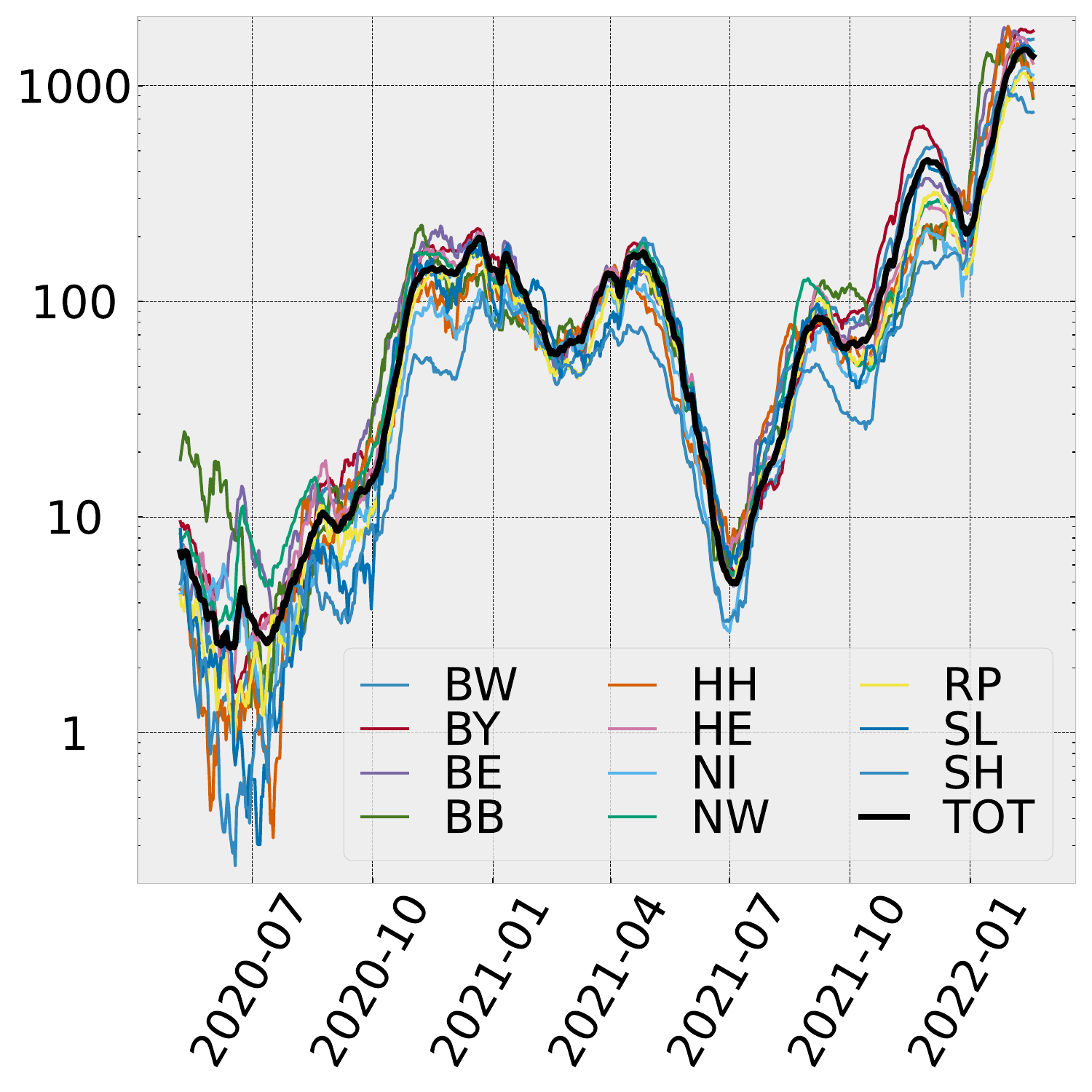} \,
      \includegraphics[height=0.52\textwidth]{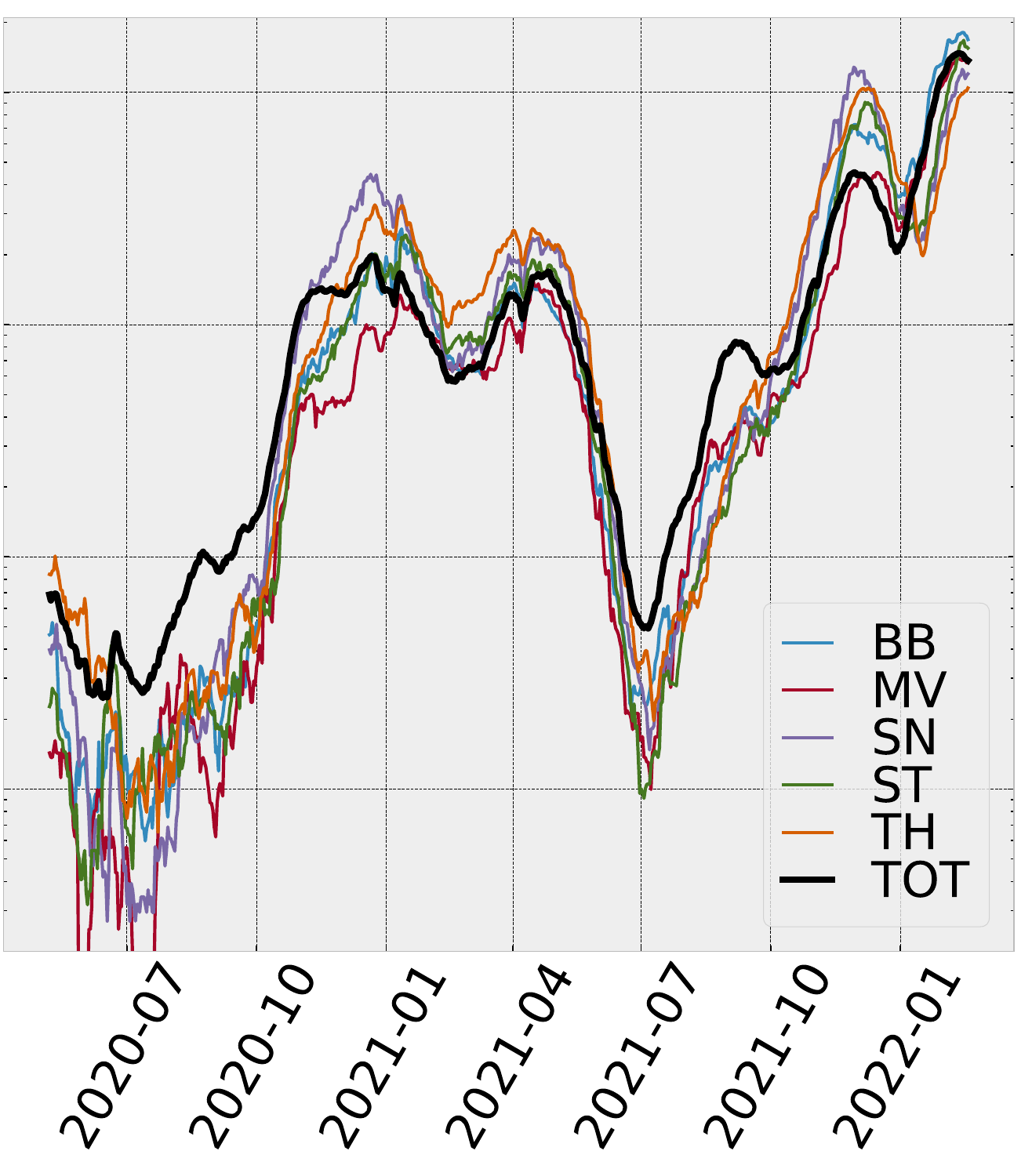}
     \caption{\textbf{COVID-19 incidence proxy} (weekly PCR-positive cases per $10^5$ inhabitants, log-scale) of  Western and Eastern German Federal States, as listed by the Robert Koch Institute (the German CDC). While  confounded by fluctuating test frequencies, the heterogeneous pattern of more and less synchronized phases indicates imperfect self-averaging, and conspicuous turning points (21-07 delta, 22-01 omicron) hint at substantial genetic-drift contributions.} 
    \label{fig:incidence}
\end{figure}

Figure~\ref{fig:incidence} displays normalized German COVID case numbers to wrap up and illustrate some of these points. The dynamics exhibits large irregular fluctuations, featuring substantial upward and downward excursions, as well as considerable spatial scatter. That such non-trivial heterogeneities persist on the spatial scales of larger parts of a densely populated country like Germany, and on time scales of years, is suggestive of poor spatiotemporal self-averaging. There appear to be extended phases of close temporal synchronization, during which the averaged incidences exhibit exponential growth and decay, at diverse rates. Whether and how these patterns emerge from the local case numbers may be contingent on the  ``art of averaging''~\cite{rudiger-etal:2021,diekmann:2000}, though. One may also perceive an underlying long-term trend of exponentially increasing infection numbers. If real, it could be indicative of a ``directed'' evolution of the rates and affinities of the pathogen-host chemistry by antigenic drift and selection. We may then be witnessing an evolutionary self-selection of increasingly contagious variants, acting as drivers of the repeated infection waves via repeated mutational bursts. And a linear net increase of effective binding energies would then account for a trend of exponentially increasing infection numbers, across the train of repeated reinfection waves. 

Other factors, such as seasonal effects, information feedback, and state regulations, should certainly not light-handedly be dismissed. Above, I have stressed particularly the role of spreading information about the course of the epidemic. This  feedback effect  is intricately intertwined, on yet another superimposed level, with the already complex spatiotemporal and genetic spreading dynamics. Game theory hints at the timely feedback and public sharing of accurately contextualized information as key elements of a potential counter-tactic that harnesses swarm intelligence to mitigate the impact of an epidemic. However, just in the midst of the information era, there appear to be technological, political, economic, and social trends that threaten to cut down on this genuinely human strength.  We must fear to forgo the potential benefits that could be reaped by bringing our swarm intelligence to bear against the swarm algorithms executed by invading pathogens. Which prompts me to close this chapter by recalling  D.~A.~Henderson's\footnote{Epidemiologist and Dean of the Johns Hopkins School of Public Health, who directed the program credited for eradicating smallpox throughout the world} ``overriding principle'' of epidemiology, namely that ``experience has shown that communities faced with epidemics or other adverse events respond best and with the least anxiety when the normal social functioning of the community is least disrupted''\cite{BiosecurityDiseaseMitigationMeasures}, and  C.~G.~Jung's insistent warnings that there is no adequate protection against psychic epidemics, which are infinitely more devastating than the worst of natural catastrophes. 


{\small
\setlength{\bibsep}{0.0pt}

}
\end{document}